\documentclass[twocolumn,aps,prx,nofootinbib]{revtex4-2}
\usepackage{graphicx}
\usepackage{bm}
\usepackage[linktocpage=true,colorlinks=true,urlcolor=blue,linkcolor=red,citecolor = blue]{hyperref}
\usepackage{pgfplots}
\usepackage{amsfonts}
\usepackage{amsmath}
\usepackage{amssymb}
\usepackage{xcolor}
\usepackage{braket}
\usepackage{relsize}
\usetikzlibrary{arrows}
\usepackage{enumerate}
\usepackage{pgfplots}
\usepackage{newfloat}
\usepackage{siunitx}
\usepackage{physics}
\usepackage{mathtools}
\usepackage{comment}
\usepackage{soul}
\usepackage[T1]{fontenc}
\usepackage{nicematrix}
\usepackage{times}

\newcommand{\printfnsymbol}[1]{%
  \textsuperscript{\@fnsymbol{#1}}%
}

\newcommand{\BS}[1]{\boldsymbol{#1}}
\newcommand{\T}[1]{\text{#1}}

\newcommand{\plaquette}[4]{%
  \begin{tikzpicture}[scale=0.3, baseline=1.3 pt]
    \draw[thick] (0,0) rectangle (1,1);
    \fill[#1] (0.25,0.75) circle(0.1);
    \fill[#2] (0.75,0.75) circle(0.1);
    \fill[#3] (0.25,0.25) circle(0.1);
    \fill[#4] (0.75,0.25) circle(0.1);
  \end{tikzpicture}%
}

\DeclareFloatingEnvironment[name={Supplementary Figure}]{suppfigure}

\graphicspath{ {Figures2/} }

\begin{document}

\title{Negative exchange interaction in Si quantum dot arrays via valley-phase induced $\mathbb{Z}_2$ gauge field}
\author{Benjamin D. Woods}
\affiliation{Department of Physics, University of Wisconsin-Madison, Madison, WI 53706, USA}

\begin{abstract}
The exchange interaction $J$ offers a powerful tool for quantum computation based on semiconductor spin qubits. 
However, the exchange interaction in two-electron systems in the absence of a magnetic field is usually constrained to be non-negative $J \geq 0$, which inhibits the construction of various dynamically corrected exchange-based gates.
In this work, we show that negative exchange $J < 0$ can be realized in two-electron Si quantum dot arrays in the absence of a magnetic field due to the presence of the valley degree of freedom.
Here, valley phase differences between dots produce a non-trivial $\mathbb{Z}_2$ gauge field in the low-energy effective theory, which in turn can lead to a negative exchange interaction.
In addition, we show that this $\mathbb{Z}_2$ gauge field can break Nagaoka ferromagnetism and be engineered by altering the occupancy of the dot array. 
Therefore, our work uncovers new tools for exchange-based quantum computing and a novel setting for studying quantum magnetism.


\end{abstract}

\maketitle

\section{Introduction}

Gate-defined semiconductor quantum dots represent a promising platform for quantum computation \cite{Loss1998,Kloeffel2013,Chatterjee2021,Stano2022} and quantum simulation \cite{Barthelemy2013,Hensgens2017,Kim2022,Wang2023}, where recent experiments \cite{Mills2021,Xue2022,Noiri2022} have demonstrated single and two-qubit gates exceeding the error correction threshold \cite{Fowler2012}. 
While there are numerous types of quantum dot qubits \cite{Burkard2023}, the great majority make use of the exchange interaction between electrons. 
In its textbook form in which two electrons occupy a double quantum dot, the exchange interaction results in the lowering of the spin-singlet relative to the spin-triplets, as the Pauli exclusion principle disallows a triplet from doubly occupying the lowest-energy orbital of a single dot \cite{Burkard1999,Burkard2023}.
As some principle advantages for quantum computation, the exchange interaction is controlled all-electrically, allows for Pauli spin blockade initialization and readout \cite{Petta2005,Maune2012}, and enables universal quantum computation with only baseband voltage pulses \cite{Levy2002,Weinstein2023}.

A key disadvantageous property of the exchange interaction between electrons in neighboring dots is that its typically constrained to be non-negative (where we use the convention that $J > 0$ energetically favors an antiferromagnetic ordering of spins).
Indeed, there exists a two-electron ground state theorem (TEGST) often quoted in the quantum dot literature that the ground state of a two-electron system under certain assumptions is guaranteed to be a spin singlet \cite{Lieb1962}. 
This constraint disallows various types of dynamically corrected gates that rely on a change of the sign in the Hamiltonian parameters to decouple the qubit system from environmental noise \cite{Khodjasteh2009,Wang2012,Kestner2013,Wang2014}.
In order to avoid this constraint, sidestep the TEGST, and achieve a negative exchange interaction $J < 0$, previous works have considered dots with higher-electron occupancy \cite{Lindemann2002,Martins2017,Deng2018b,Malinowski2018,Deng2020} or placing the dots in a significant out-of-plane magnetic field \cite{Wagner1992,Burkard1999,Zumbuhl2004,Baruffa2010,Mehl2014}.
However, large out-of-plane magnetic fields are impractical for spin qubit operation and high-electron occupancy can lead to a complicated many-body spectrum.

In this work, we show that a negative exchange interaction can be realized in a two-electron Si quantum dot system in the absence of a magnetic field.
Here, the realization of negative exchange and avoiding the TEGST relies on the presence of the valley degree of freedom.
Specifically, we show that valley phase differences between dots leads to an effective $\mathbb{Z}_2$ gauge field, defined as the signs ($\pm 1$) of the effective hopping amplitudes between dots in the low-energy theory. 
Negative exchange interactions are then realized in quantum dot plaquettes with the combination of an odd number of $+1$ gauge fields and two-electron occupancy.
Such plaquettes are characterized by a gauge-invariant $\pi$-flux that is equivalent to a (superconducting) magnetic flux quantum $\Phi_0$ threading through the plaquette. 
We stress that such realizations rely upon the two-dimensional nature of the quantum dot array, as the formation of quantum dot loops is essential for defining the gauge-invariant flux.
Therefore, our proposal takes advantage of the recent fabrication advances that extend quantum dot arrays into a second dimension \cite{Ha2021,Unseld2023,Acuna2024,Borsoi2024,Zhang2024,Wang2024,Wang2024b,Ha2025}.

\begin{figure}[b]
\begin{center}
\includegraphics[width=0.48\textwidth]{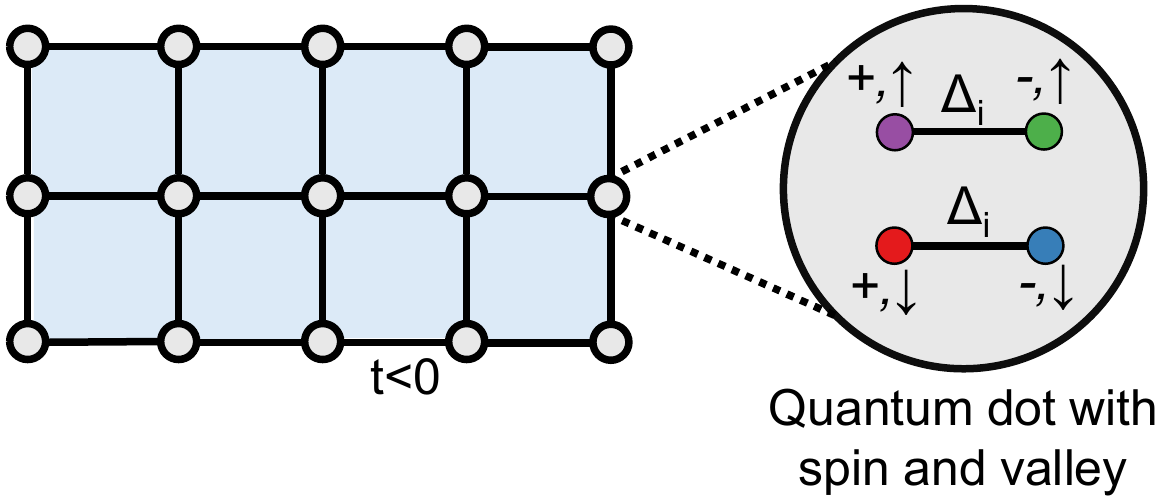}
\end{center}
\vspace{-0.5cm}
\caption{A quantum dot array, where each dot has both spin $\sigma \!\in\! \{\uparrow,\downarrow\}$ and valley $\tau \in \{+,-\}$ degrees of freedom, as indicated by the four colored circles in the blown up dot at the right. 
A black line connecting dots $i$ and $j$ denotes tunnel coupling $t_{i,j}$ that preserves spin and valley. 
Here, $t_{i,j} < 0$ due to the s-wave symmetry of each dot's lowest-energy orbital.
The valleys in dot $i$ are coupled by $\Delta_i = |\Delta_i|e^{i\phi_i}$, where $E_{v,i} = 2|\Delta_i|$ is the valley splitting and $\phi_i$ is the valley phase.
The relative valley phases between dots plays a key role in the low-energy physics.}
\label{FIGqdArray}
\vspace{-1mm}
\end{figure}

Importantly, while the $\mathbb{Z}_2$ flux configuration of a given quantum dot array is random due to the random nature of the valley phase, we show that the flux configuration can be engineered by altering the electron occupancy of selective dots.
The addition of two electrons to a dot fills its ground valley, making its ground and excited valleys \textit{inert} and \textit{active}, respectively.
As we show below, this effectively changes the valley phase of a dot by $\pm \pi$, allowing us to engineer the $\mathbb{Z}_2$ flux configuration of the quantum dot array.
In principle, this allows for the realization of negative exchange interaction in any given plaquette.
Therefore, our results offer new tools for dynamically corrected exchange-based gates in Si quantum dot arrays. 

\section{Model}\label{Model}
Consider a Si quantum dot array, as illustrated in Fig. \ref{FIGqdArray}.
The low-energy physics can be captured by a Hubbard-like Hamiltonian \cite{Buterakos2021}, where each dot has a single spatial orbital with both spin $\sigma \in \{\uparrow,\downarrow\}$ and valley $\tau \in \{+,-\}$ degrees of freedom.
Here, the valley degree of freedom comes from the existence of two degenerate valleys in the Si band structure near the $Z$ point of the Brillouin zone \cite{Zwanenburg2013,Gyure2021}.
Explicitly, the Hamiltonian is given by
\begin{equation}
    \begin{split}
    H =& \sum_{i} \varepsilon_{i} \hat{n}_i 
    + \sum_{i}\sum_{\sigma} \left( 
    \Delta_i c_{i,-,\sigma}^\dagger c_{i,+,\sigma} + h.c.
    \right) \\
    &+ \sum_{\left<i,j\right>} \sum_{\tau,\sigma} t_{i,j} c_{i,\tau,\sigma}^\dagger c_{j,\tau,\sigma}
    + \frac{U}{2} \sum_{i} \hat{n}_i\left(\hat{n}_i - 1\right) \\
    &+ \frac{1}{2}\sum_{i,j}^{i\neq j} V_{i,j} \hat{n}_i\hat{n}_j, 
    \end{split} \label{H1}
\end{equation}
where $c_{i,\tau,\sigma}^\dagger$ creates an electron with valley $\tau$ and spin $\sigma$ in dot $i$, and $\hat{n}_{i} = \sum_{\tau,\sigma} c_{i,\tau,\sigma}^\dagger c_{i,\tau, \sigma}$ is the number operator.
Here, $\varepsilon_{i}$ represents the dot potentials, while $t_{i,j} = t_{j,i}$ denotes the inter-dot tunnel couplings, both of which are tunable via gate voltages.
Importantly, the tunnel couplings are negative $t_{i,j} \leq 0$ due to the s-wave symmetry of the lowest-energy orbital of each dot.
$U$ is the Hubbard onsite Coulomb energy, which penalizes double occupancy of a dot, and $V_{i,j}$ is the inter-dot (screened) Coulomb energy.
$\Delta_i = |\Delta_i|e^{i\phi_i}$ is the complex valley coupling of dot $i$, where $E_{v,i} = 2|\Delta_i|$ is the valley \textit{splitting} and $\phi_i \in (-\pi,\pi]$ denotes the valley \textit{phase}. 
Importantly, both $E_{v,i}$ and $\phi_{i}$ vary from dot to dot, primarily due to alloy disorder fluctuations \cite{Wuetz2021,Losert2023,McJunkin2021,Lima2023a,Lima2023b}.
Indeed, in the so-called disordered regime, where alloy disorder fluctuations dominate over deterministic contributions, the valley phase is uniformly distributed and essentially uncorrelated between any given 2 dots.
Finally, note that the inter-dot tunnel couplings $t_{i,j}$ preserve both the spin and valley degrees of freedom. 
Therefore, our model is neglecting effects from spin-orbit coupling and and inter-dot inter-valley coupling, which are both expected to be small.  

\begin{figure}[t]
\begin{center}
\includegraphics[width=0.48\textwidth]{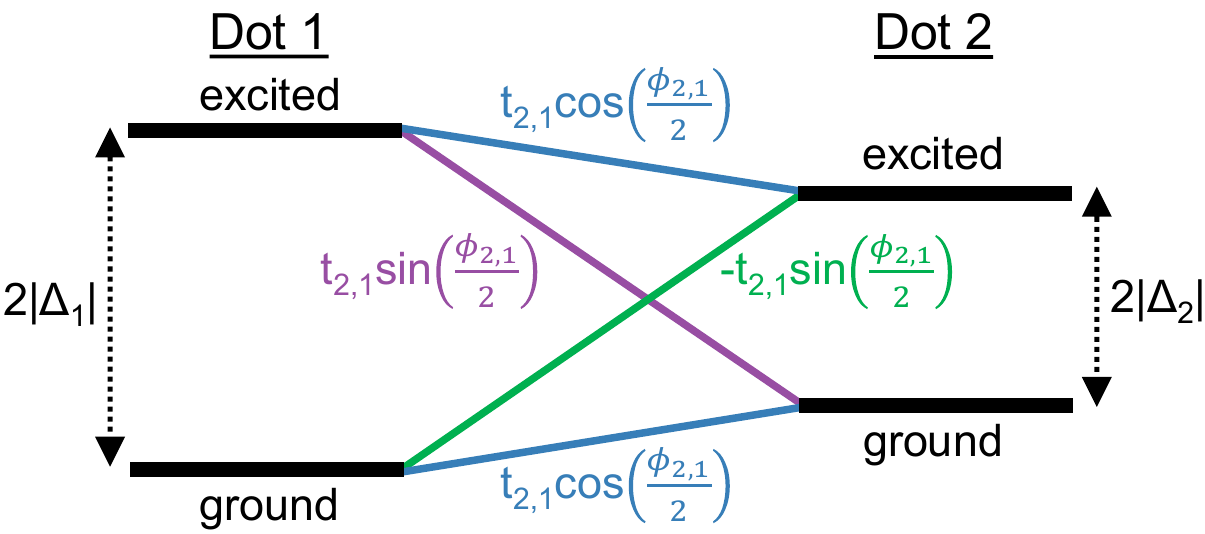}
\end{center}
\vspace{-0.5cm}
\caption{Energy level diagram of two coupled dots after transforming into the ground and excited valley basis. The ground and excited valleys of each dot are separated in energy by the valley splitting $E_{v,i} = 2|\Delta_i|$. There exists both intra-valley and inter-valley tunnel couplings with both the magnitudes and signs being determined by the valley phase difference $\phi_{2,1} = \phi_2 - \phi_1$ between the dots.}
\label{FIGlevelDiagram}
\vspace{-1mm}
\end{figure}

The relative valley phases between the dots play a key role in understanding the low-energy physics of the system.
This becomes most evident by transforming the Hamiltonian from the $\{+,-\}$-valley basis into the ground and excited valley basis, which diagonalizes the valley coupling $\Delta_i$ terms in Eq. (\ref{H1}).
We define new creation operators $\tilde{c}_{i,\tau,\sigma}^\dagger = \sum_{\tau^\prime} c_{i,\tau^\prime,\sigma}^\dagger \mathcal{U}_{\tau^\prime,\tau}(\phi_i)$, where $\tau \in \{g,e\}$ denotes the ground and excited valley, respectively, and $\mathcal{U}(\phi_i)$ is a $\phi_i$-dependent unitary matrix given in the Appendix \ref{AppA}.
The Hamiltonian in Eq. (\ref{H1}) can then be rewritten as
\begin{equation}
    \begin{split}
        H =& \sum_{i} \BS{\tilde{c}}_{i}^\dagger \big(
        \varepsilon_{i} - \tau_z |\Delta_i|
        \big) \BS{\tilde{c}}_i 
        + \sum_{\left<i,j\right>}
        \BS{\tilde{c}}_{i}^\dagger
        \tilde{t}_{i,j}(\phi_{i,j})
        \BS{\tilde{c}}_{j} \\
        &+ \frac{U}{2} \sum_{i} \hat{n}_i\left(\hat{n}_i - 1\right) 
        + \frac{1}{2}\sum_{i,j}^{i\neq j} V_{i,j} \hat{n}_i\hat{n}_j,
    \end{split} \label{H2}
\end{equation}
where $\BS{\tilde{c}}_{i} = (\tilde{c}_{i,g,\uparrow} , \tilde{c}_{i,g,\downarrow}, \tilde{c}_{i,e,\uparrow}, \tilde{c}_{i,e,\downarrow})^T $, and $\tau_{j}$ with $j = x,y,z$ are Pauli matrices acting in $\{g,e\}$-valley space.
Here, $\tilde{t}_{i,j}$ is a tunneling matrix that depends on the relative valley phase $\phi_{i,j} = \phi_{i}\! - \phi_{j}$ between two dots and is given by 
\begin{equation}
    \tilde{t}_{i,j} = 
    \cos(\phi_{i,j}/2)
        + \sin(\phi_{i,j}/2) i\tau_y.
\end{equation}
Here, we see that there exists both intra-valley and inter-valley tunnel coupling between dots, with their relative magnitudes and signs depending on the valley phase differences, as illustrated in Fig. \ref{FIGlevelDiagram}, which shows the single-particle energy level diagram of two dots in the ground and excited valley basis.
In the extreme case of $\phi_{i,j} = 0$, the inter-valley tunnel coupling is extinguished. 
In the opposite extreme of $\phi_{i,j} = \pi$, the valleys interchange character between the two dots, and the intra-valley tunnel coupling vanishes.
Note that the transformed tunnel couplings are all real.

\subsection{Effective Hamiltonian with $\mathbb{Z}_2$ gauge field}
\begin{figure}[t]
\begin{center}
\includegraphics[width=0.48\textwidth]{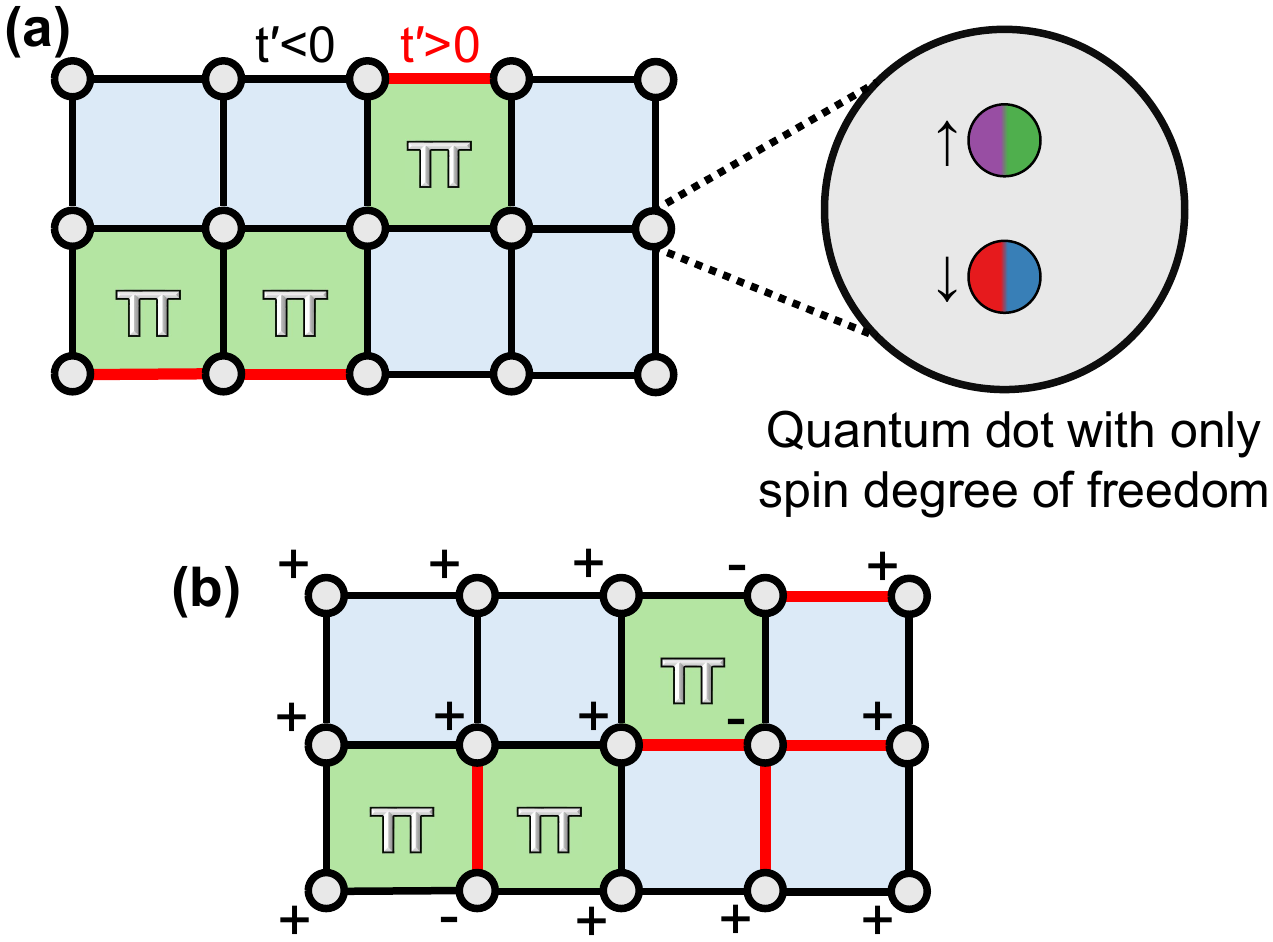}
\end{center}
\vspace{-0.5cm}
\caption{(a) Effective quantum dot array after projecting onto the ground valley of each dot. Each dot only a spin degree of freedom, as indicated by the two circles in the blown up dot. Here, the two colors of each circle indicates that the ground valley is an equal superpositions of the + and - valleys shown in Fig. \ref{FIGqdArray}. The sign of the effective tunnel coupling $t^\prime_{i,j}$ is determined by the valley phase difference $\phi_{i,j} = \phi_{i} - \phi_{j}$, where $t_{i,j}^\prime > 0$ ($t_{i,j}^\prime< 0$) are indicated by red (black) lines. 
The sign of the effective tunnel coupling defines a $\mathbb{Z}_2$ gauge field on each link between dots, $\chi_{i,j} = \T{sgn}(t_{i,j}^\prime) = \pm 1$.
Plaquettes with an odd number of $t_{i,j}^\prime > 0$ tunnel couplings have a gauge-invariant $\pi$-flux, which is equivalent to a (superconducting) magnetic flux quantum $\Phi_0$ threading through the plaquette. Such $\pi$-fluxes can lead to interesting phenomena, such as negative exchange interactions and broken Nagaoka ferromagnetism, as demonstrated below.
(b) System in (a) after performing the gauge transformation indicated by the $\pm$ factors near each dot.
Notice that the $\mathbb{Z}_2$ gauge field configuration changes, but the $\mathbb{Z}_2$ flux configuration is invariant under a gauge transformation.
}
\label{FIGgvArray}
\vspace{-1mm}
\end{figure}

Remarkably, the effects from the relative valley phases can lead to a low-energy Hamiltonian with a $\mathbb{Z}_2$ gauge field.
To see this, consider $N$ quantum dots with $M < N$ electrons. 
For simplicity, let us first consider vanishing extended Coulomb interactions, $V_{i,j} = 0$.
We consider the effects of $V_{i,j} \neq 0$ later. 
In the limit of large onsite Coulomb energy $U$, the low-energy subspace consists of states without double occupied dots. 
Furthermore, if the valley splittings dominate over the inter-dot detunings and tunnel couplings, $|\Delta_i| \gg |\varepsilon_{j}^\prime - \varepsilon_{k}^\prime|, |t_{j,k}|$, where $\varepsilon_{i}^\prime = \varepsilon_{i} - |\Delta_i|$ is a ground valley energy, then occupation is restricted to the ground valley in the low-energy subspace.
A simple truncation of the Hilbert space to the low-energy subspace described above yields the effective ground-valley Hamiltonian
\begin{equation}
    H_{eff}^{\T{g.v.}} = P
    \left[
    \sum_{i,j,\sigma} 
    \left(\delta_{i,j} \varepsilon_{i}^\prime + t_{i,j}^\prime\right) 
    \tilde{c}_{i,g,\sigma}^\dagger 
    \tilde{c}_{j,g,\sigma}
    \right] P, \label{Hgv}
\end{equation}
where $\varepsilon_{i}^\prime = \varepsilon_{i} - |\Delta_i|$ are the effective potentials, $t_{i,j}^\prime = t_{i,j} \cos(\phi_{i,j}/2)$ are effective tunnel couplings, and $P$ is a projection operator that excludes double occupancy of any dots and excited-valley occupation.
Here, the $\mathbb{Z}_2$ gauge field $\chi_{i,j}$ is defined on the links between dots and is determined on each link by the sign of its effective tunnel coupling $t_{i,j}^\prime$, $\chi_{i,j} = \T{sgn}(t_{i,j}) = \pm 1$.
In the absence of valley phase differences (i.e. $\phi_{i,j} = 0$ for all $i,j$), all effective tunnel couplings would be non-positive, $t_{i,j}^\prime \leq 0$, corresponding to a trivially uniform gauge field, $\chi_{i,j} = -1$.
However, some of the effective tunnel couplings can flip sign due to valley phase differences, where $t_{i,j}^\prime \geq 0$ whenever $|\phi_{i,j}| > \pi$.
A schematic example of this is shown in Fig. \ref{FIGgvArray}, where a red line indicates $t_{i,j}^\prime \geq 0$ and 
$\chi_{i,j} = 1$.

A physically important flux can be defined on any given plaquette as the product of all the gauge fields on the perimeter of the plaquette. 
Indeed, a plaquette with an odd number of $\chi_{i,j} = 1$ is said to be threaded by a $\pi$-flux, as indicated by the green shading in Fig. \ref{FIGgvArray}. 
Here, the name $\pi$-flux comes from the connection with the total Aharonov-Bohm phase accumulated around a plaquette that is threaded by a (superconducting) magnetic flux quantum $\Phi_0 = \pi \hbar/e$.
Note that a $\pi$-phase is precisely the phase needed to flip the sign of one tunnel coupling $t_{i,j}^\prime$ along the perimeter of the plaquette. 
Importantly, the flux of a plaquette is invariant under a gauge transformation, unlike the $\mathbb{Z}_2$ gauge field.
Fig. \ref{FIGgvArray}(b) shows the system in Fig. \ref{FIGgvArray}(a) after the gauge transformation indicated by the $\pm$ factors near each dot in Fig. \ref{FIGgvArray}(b).
We see that while the links for which $t_{i,j}^\prime > 0$ ($\chi_{i,j} = 1$) have changed, the configuration of $\pi$-fluxes has remained unchanged. 

We stress that a $\pi$-flux is a non-trivial effect arising from the valley phase differences between dots along the perimeter of a plaquette. 
Furthermore, the system needs to be hole-doped away from $1$ electron per dot (i.e. $M < N$) in order for the $\pi$-flux configuration to make an impact.\footnote{All states with $1$ electron per dot (i.e. $M = N$) are trivial eigenstates of the effective Hamiltonian in Eq. (\ref{Hgv}) since double occupancy is not allowed and our projection ignores the exchange proportional to $U^{-1}$.}  

\section{Plaquettes threaded by an effective $\pi$-flux}
We now illustrate how a $\pi$-flux can impact the low-energy physics of a few example systems.
As we show below, these $\pi$-fluxes can lead to negative exchange interactions and also destroy Nagaoka ferromagnetism. 
In this section, we assume the projection in Eq. (\ref{Hgv}) onto the ground-valley subspace accurately captures the low-energy physics.
We will discuss a situation in which this approximation breaks down later in Sec. \ref{LVS}.

\subsection{Triangular plaquette}
\begin{figure}[t]
\begin{center}
\includegraphics[width=0.48\textwidth]{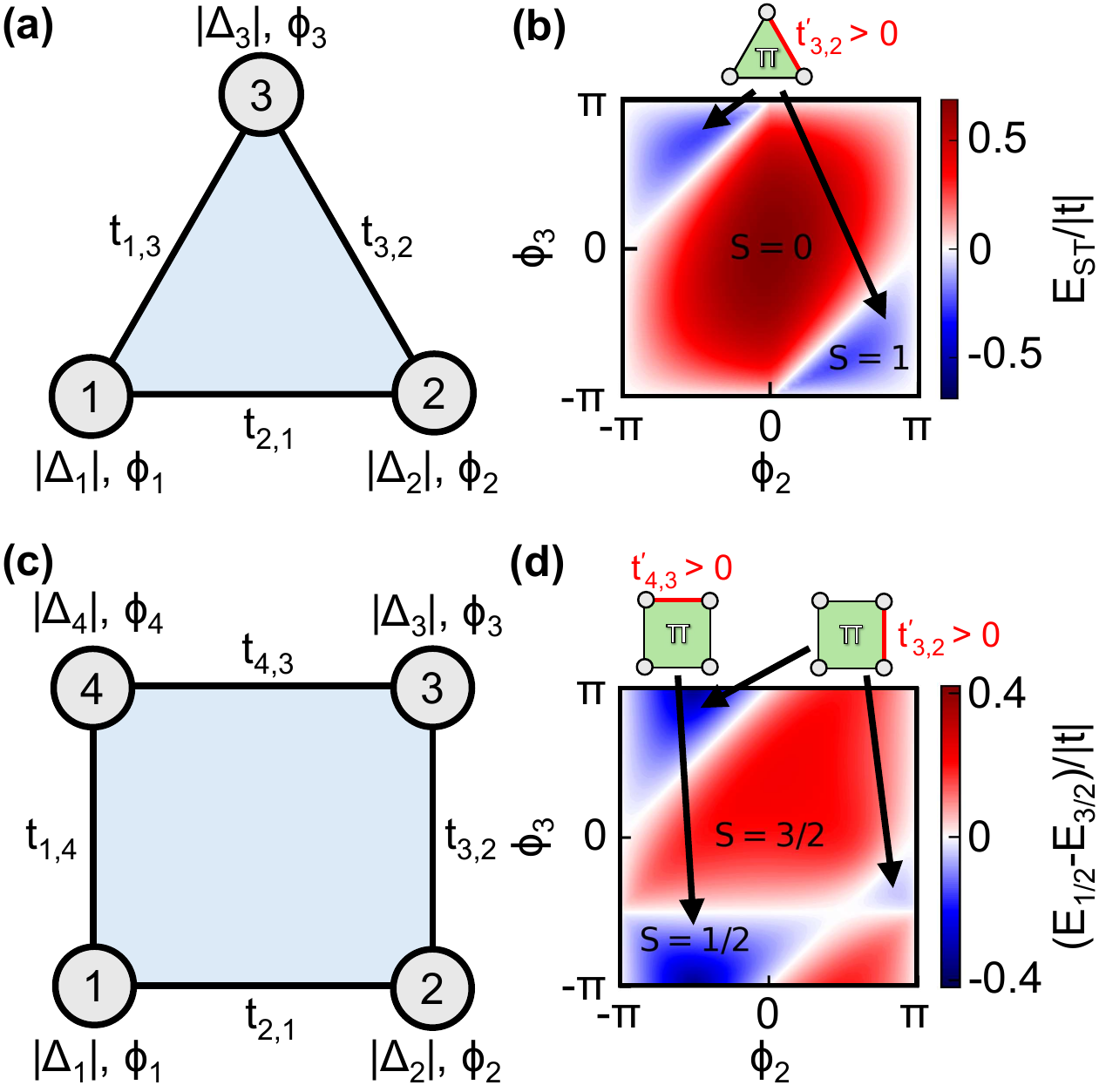}
\end{center}
\vspace{-0.5cm}
\caption{(a) Triangular quantum dot plaquette. Each dot has an inter-valley coupling $\Delta_i = |\Delta_{i}|e^{i\phi_i}$, where $\phi_{i}$ is the valley phase. Solid black lines indicate inter-dot tunnel couplings $t_{i,j} \leq 0$. (b) Singlet-triplet splitting $E_{\T{ST}}$ for $M = 2$ electrons in a triangular plaquette as a function of $\phi_2$ and $\phi_3$. Without loss of generality, we set $\phi_1 = 0$. Other parameters are $t_{i,j} = t < 0$ for all $i,j$, $|\Delta_{i}| = 50|t|$ and $\varepsilon_{i}^\prime = \varepsilon_i - |\Delta_i| =  0$ for all $i$, and $U = 1000 |t|$. 
$E_{\T{ST}} < 0$ (i.e. a negative exchange interaction $J < 0$) is realized in the blue regions, covering $1/4$ of the valley phase parameter space. For these regions, $t_{3,2}^\prime\! > \! 0$ ($\chi_{3,2} = 1$), yielding a $\pi$-flux threading the plaquette in the low-energy theory given in Eq. (\ref{Hgv}).
(c) Square plaquette.
(d) Energy splitting between the lowest-energy $S = 1/2$ and $S = 3/2$ states for $M = 3$ electrons in a square plaquette as a function of $\phi_2$ and $\phi_3$. $\phi_1 = 0$ without loss of generality, and we set $\phi_4 = \pi/2$. Other parameters are the same as (b). 
In the absence of valley phase difference, the square plaquette exhibits Nagaoka ferromagnetism ($S = 3/2$ ground state). A $\pi$-flux breaks the Nagaoka ferromagnetism, leading to a $S = 1/2$ ground state, as demonstrated by the blue regions, where $E_{1/2} - E_{3/2} < 0$.
Note that for $M = 2$ electrons in a square plaquette, $E_{ST} < 0$ will be realized in the same regions of valley phase parameter space that have $E_{1/2} - E_{3/2} < 0$ in (d).
}
\label{FIGresults1}
\vspace{-1mm}
\end{figure}

Let us first consider $M = 2$ electrons in the triple ($N =3$) quantum dot system arranged in a triangular geometry, as illustrated in Fig. \ref{FIGresults1}(a).
We will show that a negative exchange interaction is produced by a $\pi$-flux.
Such a geometry has recently been experimentally realized in a Si/SiGe system \cite{Acuna2024}.
The $SU(2)$ symmetry of Eq. (\ref{Hgv}) (along with the parent Hamiltonian in Eq. (\ref{H1})) implies that the 2-electron sector of Eq. (\ref{Hgv}) decomposes into $1$ spin singlet and $3$ identical spin triplet blocks, with total spin angular momentum of $S = 0$ and $S = 1$, respectively.
(See Appendix \ref{SU2} for details regarding the implications of $SU(2)$ symmetry.)
The singlet and triplet blocks are found to be
\begin{equation}
    H_{S/T}^{\triangle} = \begin{pmatrix}
        \varepsilon_{1}^\prime + \varepsilon_{2}^\prime &  \pm t_{3,2}^\prime & t_{1,3}^\prime \\
         \pm t_{3,2}^\prime & \varepsilon_{1}^\prime + \varepsilon_{3}^\prime & t_{2,1}^\prime \\
        t_{1,3}^\prime & 
        t_{2,1}^\prime &
        \varepsilon_{2}^\prime + \varepsilon_{3}^\prime 
    \end{pmatrix}, \label{HST}
\end{equation}
where $+$ and $-$ correspond to the singlet and triplet blocks, respectively.\footnote{See Appendix \ref{TriApp} for the explicit definition of the singlet and triplet states. In addition, we have performed a simple gauge transformation in Eq. (\ref{HST}), as described in Appendix \ref{TriApp}, to place the $\pm$ on $t_{3,2}^\prime$ elements. Before the gauge transformation, the $\pm$ is on the $t_{1,3}^\prime$ elements.}
Without loss of generality, we can assume $\phi_{1} = 0$.\footnote{If $\phi_1 \neq 0$, we can perform a global valley rotation (i.e. the same rotation on every dot) such that $\phi_1 = 0$.}
Therefore, the tunnel couplings of dot $1$ are generically non-positive, $t_{2,1}^\prime, t_{1,3}^\prime \leq 0$, and $\chi_{2,1} = \chi_{1,3} = -1$.
In contrast, $t_{3,2}^\prime$ and $\chi_{3,2}$ can be of either sign, leading to the realization of a $\pi$-flux whenever $|\phi_{3,2}| > \pi$. 
Importantly, when $t_{3,2}^\prime \rightarrow -t_{3,2}^\prime$ the singlet and triplet Hamiltonian blocks in Eq. (\ref{HST}) interchange. Hence, whether the ground state is a singlet ($S = 0$) or triplet ($S = 1$) must also change when the sign of $t_{3,2}^\prime$ flips.
This is verified in Fig. \ref{FIGresults1}(b), where the singlet-triplet splitting $E_{ST} = E_T - E_S$ is shown as a function of $\phi_2$ and $\phi_3$ for some example parameters given in the caption. Here, $E_S$ and $E_T$ are the lowest-energy eigenvalues of the singlet and triplet blocks, respectively.
We see that a negative $E_{ST}$ (corresponding to a triplet ground state and negative exchange interaction, $J < 0$) occurs in the $1/4$ of the valley-phase parameter space in which a $\pi$-plaquette is realized.
Remarkably, $|E_{ST}|$ is on order of the bare hopping $|t|$, which is much larger than the usual exchange interaction $J = |t|^2/U$ found for a double quantum dot system at zero detuning.
This exchange interaction is dramatically larger in this triangular geometry with $M = 2$ electrons because the electrons can exchange positions while avoiding the large Coulomb energy $U$ that must be paid for the double occupation of a dot.
We also point out that the triplet ground state in the triangular plaquette with a $\pi$-flux is an instantiation of Nagaoka ferromagnetism \cite{Nagaoka1966,Tasaki1998}, as all three hoppings can be made positive by a simply gauge transformation where the basis states of dot $1$ are multiplied by $-1$.

\subsection{Square plaquette} \label{SqPlaq}
Next, let us consider $M = 3$ electrons in the $N = 4$ square plaquette shown in Fig. \ref{FIGresults1}(c). 
Such a geometry has also been realized in Si quantum dots \cite{Wang2024b}.
Here, we show that the expected Nagaoka ferromagnetism can be broken by a $\pi$-flux.
Recall that Nagaoka ferromagnetism \cite{Nagaoka1966,Tasaki1998} occurs in the $U \rightarrow \infty$ limit of single-band Hubbard models when there is one fewer electrons than the number of sites (i.e. one hole), all tunnel couplings are positive, and a connectivity condition is satisfied.
In the absence of valley degrees of freedom, these conditions are met for the square plaquette geometry,\footnote{Here, the positivity condition is satisfied because the lattice is bipartite, where the sign of the tunnel couplings can be globally flipped by a gauge transformation is which all of the orbitals on one of the sublattices is multiplied by $-1$.} and one expects an $S = 3/2$ ferromagnetic ground state when 3 electrons are present.
Indeed, such Nagaoka ferromagnetism has recently been experimentally observed in a plaquette of 4 Ge quantum dots \cite{Dehollain2020}.
When the valley physics is incorporated, however, the sign of one of the tunnel couplings within the ground-valley manifold (prior to any gauge transformation) can become positive, realizing a $\pi$-flux.
It then becomes impossible to make all tunnel couplings simultaneously positive via a unitary transformation, and the Nagaoka positivity condition is unsatisfied. 
As shown in Appendix \ref{BNPCApp}, the Nagaoka positivity condition is broken in $1/3$ of valley phase parameter space, and we expect the ground state in the $U \rightarrow \infty$ limit to have spin $S = 1/2$ instead of $S = 3/2$.
We numerically verify this by the result shown in Fig. \ref{FIGresults1}(d), where the energy splitting $\Delta E_{\frac{1}{2},\frac{3}{2}} = E_{1/2} - E_{3/2}$ between the lowest-energy $S = 1/2$ and $S = 3/2$ states is shown as a function of $\phi_2$ and $\phi_3$ for fixed $\phi_4 = \pi/2$. 
The region where the Nagaoka positivity condition is unsatisfied and a $\pi$-flux is realized perfectly coincides with $\Delta E_{\frac{1}{2},\frac{3}{2}} < 0$.

A $\pi$-flux induced by valley phase effects can also produce a negative exchange interaction $J < 0$ (i.e. triplet $S = 1$ ground state) in the case of $M = 2$ electrons in the $N = 4$ square plaquette.
At this lower filling, extended Coulomb interactions begin to play a non-trivial role, so let us reintroduce $V_{i,j} = V \neq 0$ for nearest neighbor dots.
Furthermore, we assume $V \gg |t_{i,j}|,|\Delta_i|$, such that minimization of the Coulomb energy is central in determination of the low-energy subspace.
Assuming relatively small inter-dot detunings, the low-energy charge configurations are given by $\{
\plaquette{white}{black}{black}{white},
\plaquette{black}{white}{white}{black}  
\}$, where a black dot indicates the presence of an electron.
The high-energy charge configurations are $\{ 
\plaquette{white}{white}{black}{black},
\plaquette{black}{white}{black}{white},
\plaquette{white}{black}{white}{black},
\plaquette{black}{black}{white}{white}
\}$.
If the valley splittings of each dot are large compared to the potential energy difference between the $2$ low-energy charge configurations, $(\varepsilon_{1}^\prime + \varepsilon_{3}^\prime) - (\varepsilon_{2}^\prime + \varepsilon_{4}^\prime)$, 
the relevant low-energy subspace contains states in the low-energy charge configurations with exclusively  ground valleys occupied. 
Integrating out the high-energy subspace via a second-order Schrieffer-Wolff transformation then yields the effective Hamiltonian
\begin{equation}
    H_{\T{S}/\T{T}}^{\square} = 
    \begin{pmatrix}
        \varepsilon_1^\prime + \varepsilon_3^\prime + A & C_{\pm} \\
        C_{\pm} & \varepsilon_2^\prime + \varepsilon_4^\prime +B,
    \end{pmatrix} \label{HSTSq}
\end{equation}
where $+$ and $-$ correspond to the singlet and triplet blocks, respectively, the two columns correspond to the two low-energy charge configurations $\{
\plaquette{white}{black}{black}{white},
\plaquette{black}{white}{white}{black}  
\}$, and $A$, $B$, and $C_\pm$ are second-order in the tunnel couplings.
The full expressions for $A$, $B$, and $C_{\pm}$ are given in Appendix \ref{SqApp}. 
In the simple case of $\varepsilon_{j} = 0$ for all 4 dots, we find $A = B = \sum_{i,j} t_{i,j}^2/(2V)$ and 
\begin{equation}
    C_{\pm} = \frac{2}{V}\left(t^{\prime}_{2,1}t^{\prime}_{4,3} \pm t^{\prime}_{3,2}t^{\prime}_{4,3}\right). \label{Cmain}
\end{equation}
In the trivial case in which the valley phase differences flips an even number (including 0) of tunnel couplings, the singlet is the ground state, as $|C_{+}| > |C_{-}|$. 
However, in the non-trivial case in which the sign of 1 effective tunnel coupling is flipped, yielding a $\pi$-flux, then $|C_{+}| < |C_{-}|$ and the triplet becomes the ground state (i.e. $J < 0$).
Notably, this is the same condition on the valley phase configuration that destroyed the Nagaoka ferromagnetism in Fig. \ref{FIGresults1}(d) when $M = 3$ electrons were present.
In contrast to the triangular plaquette case in Fig. \ref{FIGresults1}(b), the energy scale of the singlet-triplet splitting is no longer the bare hopping strength $\mathcal{O}(|t|)$, but is rather $\mathcal{O}(t^2/V)$.
This is due to the Coulomb penalty $V$ paid by the high-energy charge configurations that serve as intermediate virtual states between the 2 low-energy charge configurations.

\begin{figure}[t]
\begin{center}
\includegraphics[width=0.48\textwidth]{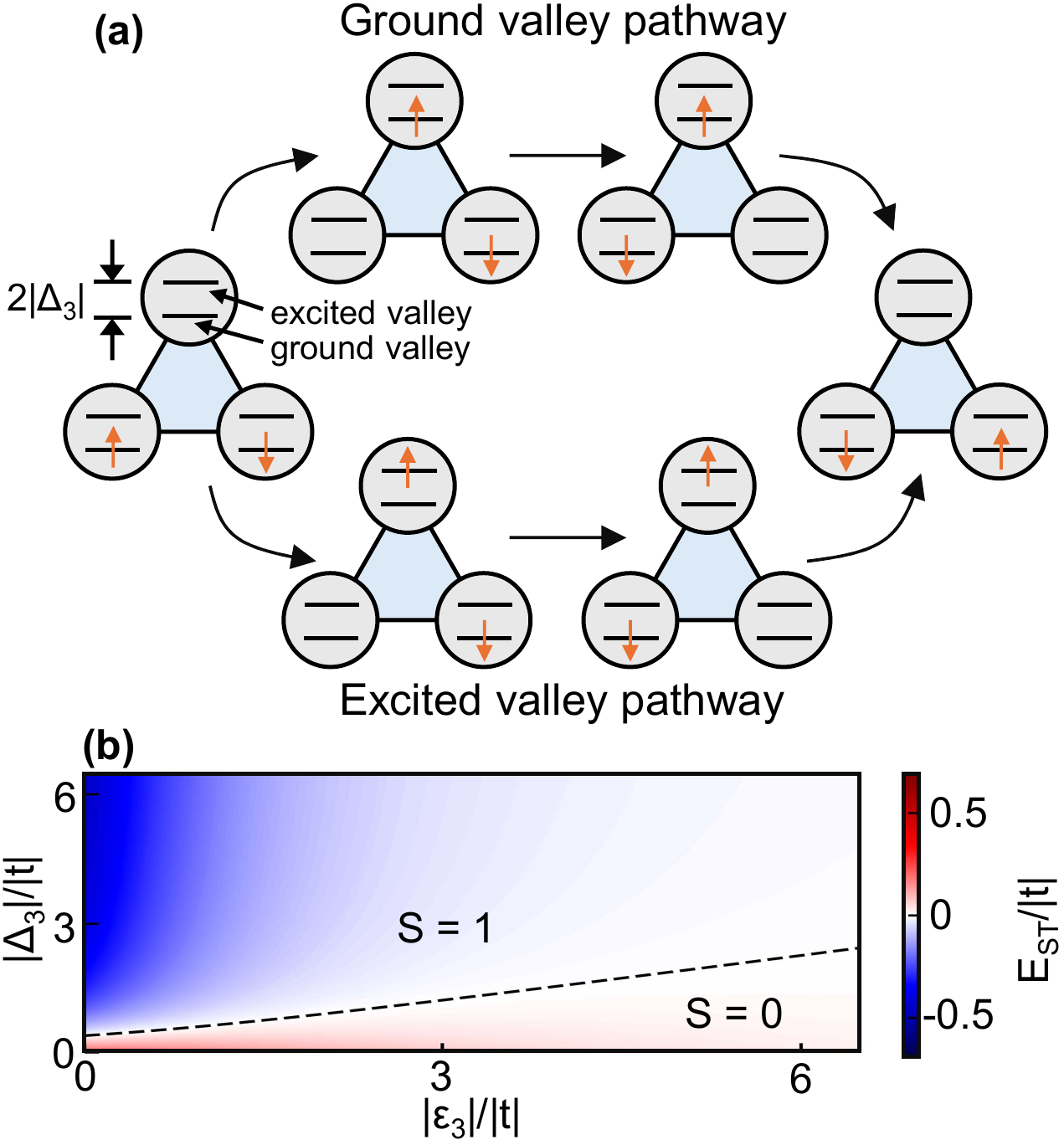}
\end{center}
\vspace{-0.5cm}
\caption{(a) Third-order processes leading to the exchange of two electrons occupying the ground valleys of the lowers dots. The upper and lower branches represent example processes involving the ground and excited valleys, respectively, of dot 3. The relative contributions of the two processes depends on the ratio of the third dot's valley splitting $2 |\Delta_3|$ and the inter-dot detuning $\varepsilon_3^\prime - \varepsilon_{1}^\prime = \varepsilon_3^\prime - \varepsilon_{2}^\prime$. (b) Singlet-triplet splitting $E_{ST}$ as a function of $\varepsilon_3^\prime$ and $|\Delta_3|$. Other parameters are $\phi_1 = 0$, $\phi_{2} = 2\pi/3$, $\phi_{3} = -2\pi/3$, $t_{i,j} = -|t| < 0$ for all $i,j$, $|\Delta_1| = |\Delta_2| = 50|t|$, and $U = 1000 |t|$. While $E_{ST} < 0$ in the limit of large $|\Delta_3|$, as consistent with Fig. \ref{FIGresults1}(b), sufficiently small $|\Delta_{3}|$ leads to $E_{ST} > 0$. 
In the $E_{ST} > 0$ region, the contribution of the third-order processes involving the excited valley of dot $3$ are counteracting and larger than the negative-exchange third-order processes involving the ground valley of dot $3$.}
\label{FIGPathways}
\vspace{-1mm}
\end{figure}

\section{Impact of low valley splitting} \label{LVS}
As a cautionary note, we point out that the various effects described above arising from valley phase differences breaks down if the valley splittings become comparable to the inter-dot detunings.
In essence, this break down occurs because the projection onto the ground valleys in Eq. (\ref{Hgv}) is unjustified.

To this see, consider again $M = 2$ electrons in the triangular $N = 3$ plaquette shown in Fig. \ref{FIGresults1}(a).
Let us assume that $\varepsilon_{1}^\prime = \varepsilon_{2}^\prime = 0$ and $\varepsilon_{3}^\prime >0$, such that the low-energy states in the absence of tunnel coupling have the ground valleys of dots $1$ and $2$ occupied.
Turning on the tunnel couplings, we see that the electrons in dots $1$ and $2$ can be exchanged by second-order and third-order processes, where the second-order processes involve an intermediate virtual state in which either dot $1$ or $2$ are doubly occupied, while the third-order processes involve virtual states with dot $3$ being occupied. 
Importantly, the intermediate virtual states are not restricted to the ground valleys.
Indeed, the third-order processes that exchange the electrons can involve either the ground valley or excited valley of dot $3$, as illustrated in Fig. \ref{FIGPathways}(a).
Therefore, the valley splittings affect the relative contributions of the various perturbation processes.
Indeed, summing over all second-order and third-order pathways (see Appendix \ref{ESTValleyApp} for full details) yields
\begin{equation}
    E_{ST} = 
    - \frac{4 t_{1,3}^\prime t_{3,2}^\prime t_{2,1}^\prime}{(\varepsilon_3^\prime)^2}
    - \frac{4 t_{1,3}^{e,g} t_{3,2}^{g,e} t_{2,1}^\prime}{(\varepsilon_3^\prime + 2|\Delta_3|)^2} 
    + J_{2,1}^{\T{direct}}, \label{EST}
\end{equation}
where $t_{i,j}^{g,e} \!= \!-t_{i,j}^{e,g}\! =\! t_{i,j}\sin(\phi_{i,j}/2)$, the first and second terms come from third-order processes involving the ground and excited valleys of dot $3$, respectively, and $J_{2,1}^{\T{direct}} \approx 4 (t_{2,1}^\prime)^2/U$ is the direct exchange between dots $1$ and $2$.
Importantly, if the first term is negative, it can be shown that the second term is guaranteed to be positive.
This implies that the perturbation processes involving the excited valley of dot $3$ counteract the negative exchange processes involving the ground valley of dot $3$.
If $\varepsilon_{3}^\prime$ is too large compared to $|\Delta_3|$, this can result in $E_{ST} > 0$ even in the region of valley parameter space where we obtain $E_{ST} < 0$ in Fig. \ref{FIGresults1}(b).
This is borne out in Fig. \ref{FIGPathways}(b), where $E_{ST}$ from an exact calculation is shown as a function of $\varepsilon_{3}^\prime$ and $|\Delta_3|$. 
Here, we see that sufficiently small $|\Delta_3|$ results in a singlet state ($S = 0$).
Therefore, we conclude that larger valley splittings are advantageous for the realization of a negative exchange interaction.

\section{Engineering $\mathbb{Z}_{2}$ flux configurations}
At first sight it appears that one has to get lucky to produce a $\pi$-flux due to the random nature of the valley phase.
However, we now show that the flux configuration can be engineered to a large degree if one allows the excited valleys of a subset of dots to be made into the active valley.
Indeed, we show that the negative exchange interaction and broken Nagaoka ferromagnetism discussed above can be realized throughout nearly the entire valley phase parameter space.
We then finally show that arbitrary configurations of $\pi$-fluxes can be realized in quasi-1D chains by valley engineering, allowing for arbitrary exchange interactions across an entire spin chain.

\begin{figure}[t]
\begin{center}
\includegraphics[width=0.48\textwidth]{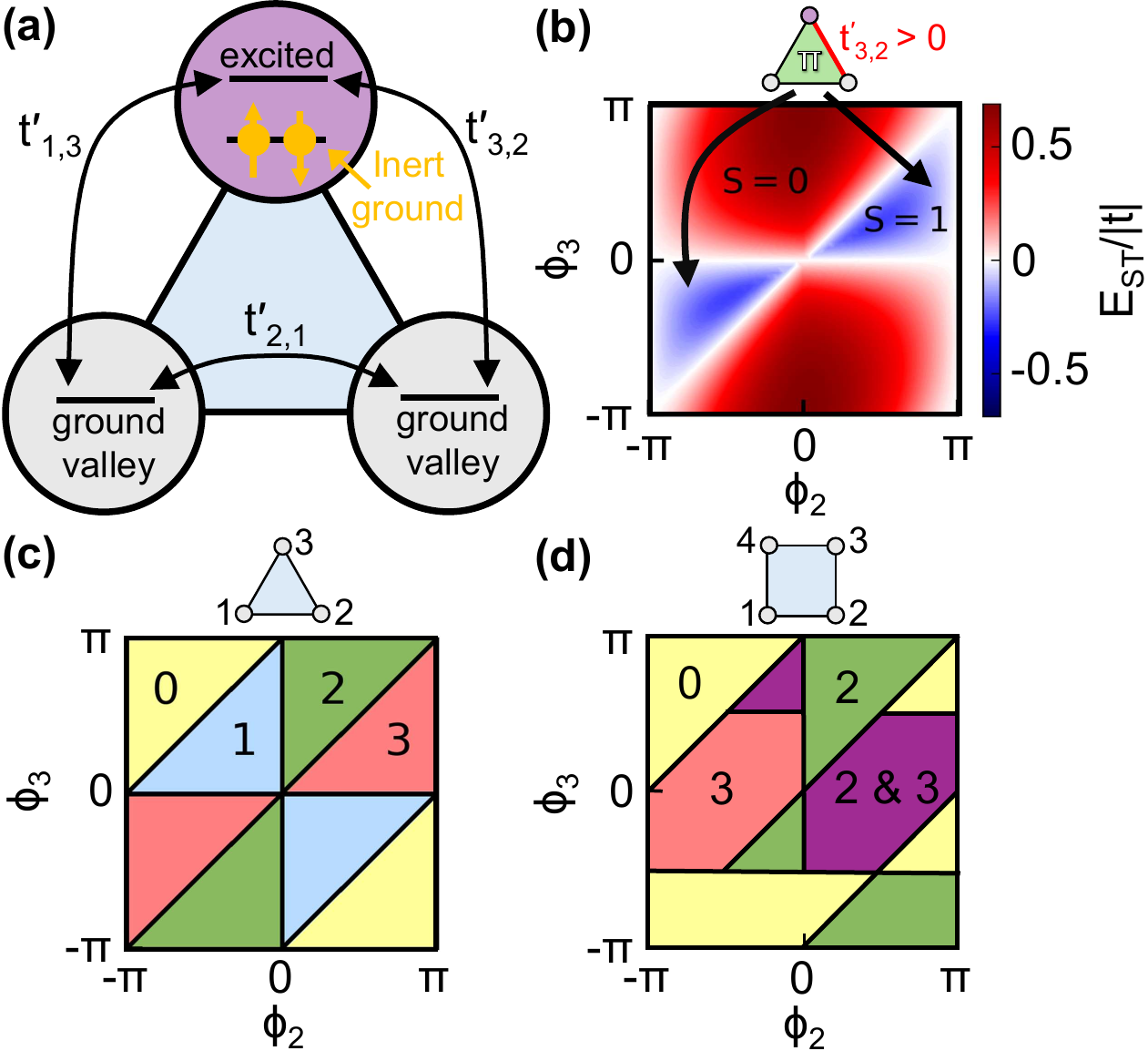}
\end{center}
\vspace{-0.5cm}
\caption{(a) Triangular plaquette in which the ground valley of dot $3$ is made inert by increasing the electron occupation to $M = 4$ electrons and sufficiently lowering $\varepsilon_3$. Here, the excited valley of dot $3$ is said to be the \textit{active} valley.
(b) Singlet-triplet splitting $E_{ST}$ for the same parameters as Fig. \ref{FIGresults1}(b), except $\varepsilon_3^\prime = \varepsilon_3 + |\Delta_3| + 3U$ is defined as the energy of the excited valley in dot $3$, such that the excited valley in dot $3$ is the active valley. 
Notice that the $\phi_3$ values of the regions with $E_{ST} < 0$ are shifted by $\pm \pi$ with respect to Fig. \ref{FIGresults1}(b).
(c) Numbers indicate that $E_{ST} < 0$ (i.e. negative exchange $J < 0$) is realized if corresponding dot has its excited valley as the active valley.
$0$ indicates all dots have their ground valleys as active valleys.
Notice that negative exchange interaction $E_{ST} < 0$ can be realized for all valley phase configurations.
(d) Same as (c), except for the case of a square plaquette with $\phi_4 = \pi/2$.
The regions labeled by $2$ \& $3$ indicates that the excited valleys of both dots $2$ and $3$ should be active.
Again $E_{ST} < 0$ (for $M = 2$ electrons) or broken Nagaoka ferromagnetism (for $M = 3$ electrons) is possible for all valley phase configurations.
}
\label{FIGActValley1}
\vspace{-1mm}
\end{figure}

To understand how negative exchange can be realized for any collection of valley phases, consider again the $N = 3$ triangular plaquette shown in Fig. \ref{FIGresults1}(a), but now with $M = 4$ electrons.
If we sufficiently lower $\varepsilon_{3}$, the ground valley of dot $3$ will be filled by two electrons for all the low-energy basis states.
The ground valley of dot $3$ is then inert, as shown in Fig. \ref{FIGActValley1}(a), and the 4-electron system will effectively behave as a 2-electron system with the Hamiltonian given in Eq. (\ref{Hgv}).
The only difference is that $\varepsilon_3^\prime = \varepsilon_{3} + |\Delta_3| + 3U$ is the energy of the excited valley in dot $3$ and $t_{i,3}^\prime = t_{i,3} \sin(\phi_{i,3}/2)$ is the effective tunneling involving dot $3$.
To understand how this affects the $\mathbb{Z}_2$ gauge field and flux configuration, consider the identity
\begin{equation}
    t_{i,3}^\prime = t_{i,3}\sin(\phi_{i,3}/2) = t_{i,3} \cos((\phi_{i} - \phi_{3} - \pi)/2).
\end{equation}
We see that making the excited valley in dot $3$ the \textit{active} valley is equivalent to $\phi_3 \rightarrow \phi_3 \pm \pi$.\footnote{We add or subtract such that $\phi_3 \! \in \! (-\pi,\pi]$. If $\phi_i^\prime\! =\!\phi_i\! +\! \pi \! >\! \pi$, then we can bring $\phi_i^\prime$ back into the range $(-\pi,\pi]$ by subtracting $-2\pi$. This has the effect of causing all hoppings involving dot $i$ to flip sign, because $\cos((\phi_i + 2\pi - \phi_j)/2) = -\cos((\phi_i - \phi_j)/2)$ and $\sin((\phi_i + 2\pi - \phi_j)/2) = -\sin((\phi_i - \phi_j)/2)$. This minus sign can be removed, however, by simply multiplying the basis states of dot $i$ by $-1$.} 
The singlet-triplet splitting $E_{ST}$ for this case is shown in Fig. \ref{FIGActValley1}(b), where all the parameters are the same as Fig. \ref{FIGresults1}(b), except the excited valley in dot $3$ is made the active valley by sufficiently lowering $\varepsilon_3$.
The regions of $E_{ST} < 0$ can be seen to be shifted by $\pm \pi$ in $\phi_3$ when compared to Fig. \ref{FIGresults1}(b), as expected from the above considerations.
This exercise can be repeated for the excited valley of dot $1$ or $2$ being made the active valley. It results in covering the entire valley phase parameter space with regions of $E_{ST} < 0$, as shown in \ref{FIGActValley1}(c). 
Therefore, we conclude that it is always possible (in the large valley splitting regime) to realize negative exchange  for an isolated triangular plaquette.

Similar considerations apply to the $N = 4$ square plaquette, where a $\pi$-flux is found to always be realizable (in the large valley splitting regime) regardless of the valley phase configuration by making an excited valley the active valley in a subset of dots.
Specifically, we that we can shift the $S = 1/2$ ground state regions of the $M = 3$ electron case in Fig. \ref{FIGresults1}(d) to any arbitrary point in the $(\phi_2,\phi_3)$-plane by making the excited valley the active valley in either dot $2$, $3$, or both $2$ and $3$. 
Indeed, Fig. \ref{FIGActValley1}(d) shows which combinations of dots $2$ and $3$ having active excited valleys realizes an $S = 1/2$ ground state for $M = 3$ across the $(\phi_2,\phi_3)$-plane for fixed $\phi_4 = \pi/2$.
Fig. \ref{FIGActValley1}(d) also applies to the realization of $E_{ST} < 0$ for $M = 2$ electrons in a square plaquette.

\begin{figure}[t]
\begin{center}
\includegraphics[width=0.48\textwidth]{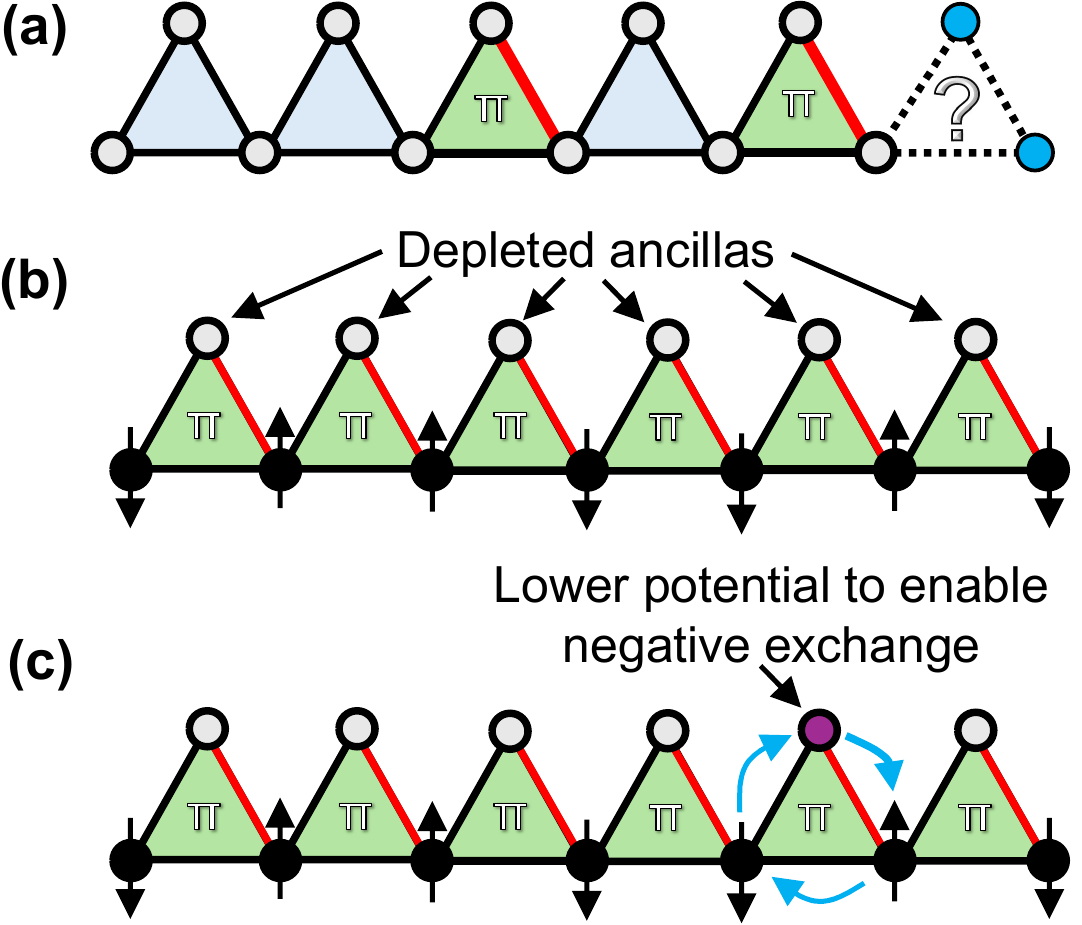}
\end{center}
\vspace{-0.5cm}
\caption{(a) Engineering the $\mathbb{Z}_2$ flux configuration of a sawtooth quantum dot array. 
An additional triangular plaquette is added onto the array by the addition of the two blue sites. We prove in the main text that a $\pi$-flux can always be engineered in the new plaquette by an appropriate choice of active valleys of the two blue sites. 
Therefore, an arbitrary flux configuration can be engineered in the sawtooth array by an appropriate choice of active valleys across the entire array.
(b) Example of a sawtooth array in which a $\pi$-flux threads all triangular plaquettes.
Here, the dot potentials are tuned such that the bottom and upper rows are occupied and depleted, respectively.
The bottom and upper rows act as computational and ancillary quantum dots, respectively. 
The exchange interaction between neighboring computational dots is positive $J_{i,j} > 0$ due to the depletion of the ancillas.
(c) The potential of an ancillary dot (shaded purple) is lowered to facilitate third-order processes (indicated by the blue arrows and shown in Fig. \ref{FIGPathways}(a)) that exchange the electrons of the computational dots.
This realizes a negative exchange interaction $J_{i,j} < 0$ when the energy of moving one electron from the two computational dots into their common ancillary dot is small compared to the valley splitting of the ancillary dot.
}
\label{FIGEngineerFlux}
\vspace{-1mm}
\end{figure}

The above results raise the question whether its possible to engineer arbitrary $\mathbb{Z}_2$ flux configurations in larger quantum dot arrays. 
While this is not always possible, we do find this is possible for several classes of arrays.
For example, consider the sawtooth chain shown in Fig. \ref{FIGEngineerFlux}(a). 
We now show that an arbitrary placement of $\pi$-fluxes can be egineered by appropriate choice of active valleys.
The proof is based on induction.
Suppose we have $Q \in \mathbb{N}_{+}$ triangular plaquettes with an arbitrary configuration of $\pi$-fluxes. 
The $Q+1$ plaquette can be appended to the edge of the system by adding $2$ additional sites, as shown as blue sites in Fig \ref{FIGEngineerFlux}(a).
The flux arrangement in the original $Q$ plaquettes is invariant under a global rotation of the valley phases.
Therefore, we can assume, without loss of generality, that the lower-left dot of the new $Q+1$ plaquette has a vanishing valley phase, $\phi = 0$.
Importantly, this is precisely the situation of the isolated $N = 3$ triangular plaquette that we have already analyzed in Fig. \ref{FIGresults1}(a) and (b). 
Furthermore, we found in Fig. \ref{FIGActValley1}(c) that a $\pi$-flux could always be realized in a triangular plaquette, independent of the valley phase configuration, by an appropriate choice of active valleys.
Clearly, we can engineer a $\pi$-flux if the valley configuration resides in the $0$, $2$, or $3$ regions of Fig. \ref{FIGActValley1}(c), as we can decide on the active valley in dots $2$ and $3$ of the new plaquette in Fig. \ref{FIGEngineerFlux}(a).
The only troubling case is if the valley configuration of the new plaquette falls in region $1$ of Fig. \ref{FIGActValley1}(c), because we cannot change the active valley of dot $1$ without affecting the previous plaquette in the chain.
Fortunately, one can show that changing the active valley in dot $1$ is equivalent to changing the active valley in both dot $2$ and $3$. 
Therefore, the $Q+1$ plaquette in our 1-dimensional array of plaquettes can always realize a $\pi$-flux if desired by an appropriate choice of the active valleys of the 2 new dots.
By induction, an arbitrary flux can be engineered for every plaquette along the 1-dimensional array.

As an application of this flux engineering, we now illustrate how negative exchange can realized on demand between any two spins along a 1-dimensional spin chain.
Such an ability may be useful in dynamical decoupling protocols for spin qubits and also engineering symmetry protected topological phases, such as the S = 1 Haldane chain \cite{Haldane1983,Shim2010,Sugimoto2020,Catarina2022,Baran2024,Manalo2024}. 
We again consider a sawtooth quantum array, as shown in Fig. \ref{FIGActValley1}(b), and assume that a $\pi$-flux has been engineered within each triangular plaquette via an appropriate choice of active valleys in the dots.
Here, the bottom row serves as the computational or active quantum dots, while the upper row provides ancillary quantum dots used for enabling the negative exchange interactions.   
Suppose that the potentials of the dots are tuned such that the computational dots in the bottom row are each occupied, while the top row of ancillary dots are depleted, as shown in Fig. \ref{FIGActValley1}(b).
In this situation, the exchange interaction between the computational dots is positive, $J_{i,j} > 0$. 
This positivity is guaranteed either by the dominance of the direct exchange interactions $J_{i,j}^{\T{direct}} > 0$ between neighboring computation dots or because the potential of the ancillary dots is large enough that the negative exchange is destroyed by the low-valley splitting mechanism discussed in Sec. \ref{LVS}.  
Negative exchange between any neighboring computational dots can then be achieved by sufficiently lowering the potential of their common ancillary dot, as shown in Fig. \ref{FIGEngineerFlux}(c).
Negative exchange will be achieved, just like in Fig. \ref{FIGresults1}(b) and Fig. \ref{FIGPathways}(b), when the energy of the moving one electron from the two computational dots to their common ancillary dot is small compared to the valley splitting of the ancillary dot.  
Note that this negative exchange interaction can be made static by parking the ancillary dot's potential at an appropriate potential or be turned on and off on demand by simply altering the ancillary dot's potential as a function of time.

\section{Conclusion}
We have shown that negative exchange interactions ($J < 0$) can be realized in two-electron Si quantum systems due to the presence of a $\mathbb{Z}_2$ gauge field arising from the valley degree of freedom. 
Hence, we have provided a counterexample to the often quoted TEGST that constrains $J > 0$ between quantum dots \cite{Lieb1962}.
Our findings may therefore be useful for performing dynamically corrected exchange-based gates that require the ability to flip the sign of $J$.
In addition, we have shown that the $\mathbb{Z}_2$ gauge field can break Nagaoka ferromagnetism and be engineered across an array by filling the ground valley of a subset of dots. Future work will study the effects of valley physics on quantum magnetism in larger quantum dot arrays.

We note that the study of systems with non-trivial $\mathbb{Z}_2$ gauge flux configurations 
has recently drawn much attention in various artificial crystals \cite{Cooper2019,Dalibard2011,Ozawa2019,Yang2015,Xue2022b,Wang2024c}, where the $\mathbb{Z}_2$ flux configuration can be engineered.
Importantly, the presence of $\pi$-fluxes alters the classification of topological phases of matter \cite{Chen2023}, leading to novel physical phenomena \cite{Xue2022c,Shao2023,Jiang2024}.
Therefore, our discovery that such $\mathbb{Z}_2$ gauge fields can be engineered via valley physics in Si quantum dot arrays opens up the possibility to study such novel topological phenomena in a new setting.
Indeed, understanding how systems with $\mathbb{Z}_2$ gauge fields are impacted by the strong Coulomb interaction that naturally occurs in quantum dot arrays may be an interesting direction for future research.

\begin{acknowledgments}
We acknowledge helpful discussions with Robert Joynt and Mark Friesen.
This work was supported part by the Army Research Office (Grant No. W911NF-23-1-0110). The views and conclusions contained in this document are those of the authors and should not be interpreted as representing the official policies, either expressed or implied, of the Army Research Office (ARO), or the U.S. Government. The U.S. Government is authorized to reproduce and distribute reprints for Government purposes notwithstanding any copyright notation herein.
\end{acknowledgments}


\appendix
\renewcommand\thefigure{\thesection.\arabic{figure}}    
\setcounter{figure}{0}

\section{Valley basis transformation} \label{AppA}
In Sec. \ref{Model} of the main text, we performed a basis transformation from the $\{+,-\}$-valley basis to the $\{g,e\}$-valley basis, where $g$ and $e$ stand for ground and excited valleys, respectively.
The $\{+,-\}$-valley basis to the $\{g,e\}$-valley basis for dot $i$ are related by 
\begin{equation}
    \tilde{c}_{i,\tau,\sigma}^\dagger = \sum_{\tau^\prime} c_{i,\tau^\prime,\sigma}^\dagger \mathcal{U}^{(i)}_{\tau^\prime,\tau}(\phi_i) \label{Trans1}
\end{equation}
where $\tau \in \{g,e\}$, $\tau^\prime \in \{+,-\}$, and $\mathcal{U}^{(i)}$ is a unitary matrix that depends on the valley phase $\phi_i$. 
In this appendix, we provide details regarding $\mathcal{U}^{(i)}$ and how it transforms the Hamiltonian.

To begin, we first note that this unitary transformation does not involve any mixing of states from different dots. 
Therefore, the number operators $n_i$ remain unaltered, implying that the interacting terms of the Hamiltonian in Eq. (\ref{H1}) of the main text are invariant.
Hence, we can focus on how the single-particle (i.e. first-quantized) Hamiltonian transforms under this unitary transformation.

The first-quantized Hamiltonian of two quantum dots in the $\{+,-\}$-valley basis is given by 
\begin{equation}
    h = \begin{pmatrix}
        A_1  & t_{2,1} \\
        t_{2,1} & A_2
    \end{pmatrix},
\end{equation}
where the first and second columns correspond to dot $1$ and dot $2$, respectively, and 
\begin{equation}
    A_i = \varepsilon_{i} + |\Delta_i|\left(\cos\phi_i \tau_x + \sin\phi_i \tau_y\right)
\end{equation}
is the intra-dot Hamiltonian in which $\tau_j$ with $j = x,y,z$ are Pauli matrices acting in valley space.
We diagonalize the valley coupling of dot $i$ by first performing a rotation by an angle $-\phi_i$ about the $\tau_z$ axis of dot $i$. This corresponds to the unitary matrix
\begin{equation}
    U_{\tau_{z}} = \begin{pmatrix}
        U_{\tau_{z}}^{(1)}(\phi_1)  & 0 \\
        0 & U_{\tau_{z}}^{(2)}(\phi_2) 
    \end{pmatrix},
\end{equation}
where 
\begin{equation}
    U_{\tau_{z}}^{(i)}(\phi_i) = \cos(\phi_{i}/2) - i\sin(\phi_{i}/2) \tau_z \label{Uz1}
\end{equation}
removes the valley phase of dot $i$.
The transformed first-quantized Hamiltonian is then found to be
\begin{equation}
    h_1 = U_{\tau_{z}}^\dagger h U_{\tau_{z}} = 
    \begin{pmatrix}
        \varepsilon_{1} + |\Delta_{1}| \tau_x & (T^{(1)})^\dagger \\
        T^{(1)} & \varepsilon_{2} + |\Delta_{2}| \tau_x
    \end{pmatrix},
\end{equation}
where $T^{(1)}$ is the tunnel coupling matrix block given by 
\begin{equation}
    T^{(1)} = 
    t_{2,1} \left[ \cos \frac{\phi_{2,1}}{2} + i \sin \frac{\phi_{2,1}}{2} \tau_z  \right],
\end{equation}
where $\phi_{2,1} = \phi_{2} - \phi_{1}$ is the valley phase difference.
Next, we perform a $\pi/2$ rotation about the $\tau_y$ axis to diagonalize the valley coupling. 
This is done with the unitary matrix
\begin{equation}
    U_{\tau_{y}} = \frac{1}{\sqrt{2}}\left( 1 + i \tau_y\right), \label{Uy}
\end{equation}
leading to the transformed Hamiltonian 
\begin{equation}
    h_2 = U_{\tau_{y}}^\dagger h_1 U_{\tau_{y}} = 
    \begin{pmatrix}
        \varepsilon_{1} - |\Delta_{1}| \tau_z & (T^{(2)})^\dagger \\
        T^{(2)} & \varepsilon_{2} - |\Delta_{2}| \tau_z
    \end{pmatrix},
\end{equation}
where $T^{(2)}$ is the tunnel coupling matrix block given by 
\begin{equation}
    T^{(2)} = 
    t_{2,1} \left[ \cos \frac{\phi_{2,1}}{2} + i \sin \frac{\phi_{2,1}}{2} \tau_x  \right],
\end{equation}
Finally, we make the inter-dot tunnel couplings purely real by rotating by $\pi/2$ about the $\tau_z$ axis.
This is done with the unitary matrix
\begin{equation}
    U_{\tau_{z},2} = \frac{1}{\sqrt{2}}\left(
    1 + i \tau_z\right), \label{Uz2}
\end{equation}
leading to the final version of the first-quantized Hamiltonian given by
\begin{equation}
    h_3 = U_{\tau_{z},2}^\dagger h_2 U_{\tau_{z},2} = 
    \begin{pmatrix}
        \varepsilon_{1} - |\Delta_{1}| \tau_z & T^\dagger \\
        T & \varepsilon_{2} - |\Delta_{2}| \tau_z
    \end{pmatrix}, \label{h3}
\end{equation}
where $T$ is the tunnel coupling matrix block given by 
\begin{equation}
    T = 
    t_{2,1} \left[ \cos \frac{\phi_{2,1}}{2} + i \sin \frac{\phi_{2,1}}{2} \tau_y  \right].
\end{equation}
The final form of $\mathcal{U}^{(i)}$ given in Eq. (\ref{Trans1}) is then
\begin{equation}
    \mathcal{U}^{(i)}(\phi_i) = U_{\tau_z}^{(i)}(\phi_i)U_{\tau_{y}} U_{\tau_{z},2},
\end{equation}
where the three factors are given in Eqs. (\ref{Uz1}, \ref{Uy}, \ref{Uz2}). This unitary transformation then results in the transformed Hamiltonian in the $\{g,e\}$-valley basis given in Eq. (\ref{H2}) of the main text.

\section{$SU(2)$ symmetry} \label{SU2}
In this appendix, we provide details regarding the consequences of the $SU(2)$ symmetry of the Hamiltonian given in Eqs. (\ref{H1}, \ref{H2}) of the main text. 
In particular, we write down basis states for $M = 2$ and $M = 3$ electron states with good angular momentum quantum numbers. 

A many-body state with $M$ electrons is defined by
\begin{widetext}
\begin{equation}
    \ket{\alpha_1 \sigma_1,\alpha_2\sigma_2,\dots,\alpha_{M-1}\sigma_{M-1},\alpha_M\sigma_M} = 
    c^{\dagger}_{\alpha_1 \sigma_1}
    c^{\dagger}_{\alpha_2 \sigma_2}\dots
    c^{\dagger}_{\alpha_{M-1} \sigma_{M-1}}
    c^{\dagger}_{\alpha_M \sigma_M}
    \ket{\T{vacuum}},
\end{equation}
\end{widetext}
where $\ket{\T{vacuum}}$ is the state with zero electrons, and $\alpha_{n} = \alpha_n(i_n,\tau_n) $ is a combined site and valley index.
Here, we impose an ordering with the convention $\alpha_{n-1} \leq \alpha_{n}$, where equality is only possible if $\sigma_{n-1},\sigma_{n} = \uparrow, \downarrow$.
Note that this ordering is important due to the anti-commutation relations of the electron creation and annihilation operators.
For $M$ electrons with $P$ single-particle states, there are in principle $P!/(M!(M-P)!)$ states.
However, these states decompose into several uncoupled sectors due to the $SU(2)$ symmetry (i.e. spin rotation symmetry) of the Hamiltonian given in Eq. (\ref{H1}) of the main text.
Indeed, defining the standard spin-operators $S_{\mu} = \frac{1}{2} \sum_{\alpha,\sigma \sigma^\prime} c_{\alpha\sigma}^\dagger (\sigma_\mu)_{\sigma \sigma^\prime} c_{\alpha \sigma^\prime}$ and $\BS{S}^2 = \sum_{\mu} S_{\mu}^2$, where $\sigma_{\mu}$ ($\mu = x,y,z$) are Pauli matrices acting in space space, we have $\left[H,S_{z}\right] = \left[H,\BS{S}^2\right] = \left[\BS{S}^2,S_z\right] = 0$. 
This implies the states can be labeled as $\ket{S,m_z,n}$, where $S$ and $m_z$ are the total and $z$-axis angular momenta, respectively, and we have the relations $\BS{S}^2 \ket{S,m_z,n} = S(S+1)\ket{S,m_z,n}$ and $S_{z}\ket{S,m_z,n} = m_z \ket{S,m_z,n}$. Following the standard treatment of quantum mechanical angular momentum, we know that for any given $S$ there are $S(S-1)$ possible $m_z$ values given by $m_z = S, S-1, \dots, -S + 1, -S$. Furthermore, the various $m_z$ states for any given $S$ are all degenerate and related by $S_{-}\ket{S,m_z,n} \propto \ket{S,m_z - 1,n}$ for $m_z \neq -S$, where $S_{-} = S_x - i S_y$ is the lowering operator.

For the case of $M = 2$ electrons, the standard quantum mechanical addition of angular momentum implies $\frac{1}{2} \bigotimes \frac{1}{2} = 0 \bigoplus 1$, i.e. there is a singlet $S = 0$ and triplet $S = 1$ sector.
The basis states within the triplet ($S = 1$) sector are given by
\begin{equation}
    \ket{T_{m_z},\alpha,\beta} = 
    \begin{cases}
        \ket{\alpha\uparrow,\beta\uparrow}, & m_z = 1 \\
        \frac{1}{\sqrt{2}}\left(\ket{\alpha\uparrow,\beta\downarrow} +  \ket{\alpha\downarrow,\beta\uparrow} \right), & m_z = 0 \\
        \ket{\alpha\downarrow,\beta\downarrow}, & m_z = -1
    \end{cases} \label{Triplets}
\end{equation}
where $\alpha < \beta$, and we remind the reader that $\alpha$ and $\beta$ are combined site and valley indices. The fact that $\alpha \neq \beta$ in the triplet sector is due to the Pauli exclusion principle.
The basis states within the singlet ($S = 0$) sector are given by 
\begin{equation}
    \ket{S,\alpha,\beta} = 
    \begin{cases}
        \frac{1}{\sqrt{2}}\left(\ket{\alpha\uparrow,\beta\downarrow}
        -  \ket{\alpha\downarrow,\beta\uparrow}
        \right), & \alpha \neq \beta \\
        \ket{\alpha\uparrow, \alpha \downarrow} & \alpha = \beta
    \end{cases}, \label{Singlet}
\end{equation}
where $\alpha \leq \beta$. In contrast to the triplet sector, $\alpha = \beta$ is allowed by the Pauli exclusion principle in the singlet sector. 

For the case of $M = 3$ electrons, the addition of angular momentum is given by $\frac{1}{2} \bigotimes \frac{1}{2} \bigotimes \frac{1}{2} = \frac{1}{2} \bigoplus \frac{1}{2} \bigoplus \frac{3}{2}$, i.e. there are two doublet $S = 1/2$ sectors and one quartic $S = 3/2$ sector. 
The basis states within the quartic ($S = 3/2$) sector are given by 
\begin{widetext}
\begin{equation}
    \ket{Q_{m_z},\alpha,\beta,\gamma} = 
    \begin{cases}
        \ket{\alpha\uparrow,\beta\uparrow,\gamma \uparrow}, & m_z = 3/2 \\
        \frac{1}{\sqrt{3}}\left(
        \ket{\alpha\downarrow,\beta\uparrow,\gamma \uparrow} 
        +  \ket{\alpha\uparrow,\beta\downarrow,\gamma \uparrow}
        + \ket{\alpha\uparrow,\beta\uparrow,\gamma \downarrow}
        \right), & m_z = 1/2 \\
        \frac{1}{\sqrt{3}}\left(
        \ket{\alpha\uparrow,\beta\downarrow,\gamma \downarrow} 
        +  \ket{\alpha\downarrow,\beta\uparrow,\gamma \downarrow}
        + \ket{\alpha\downarrow,\beta\downarrow,\gamma \uparrow}
        \right), & m_z = -1/2 \\
        \ket{\alpha\downarrow,\beta\downarrow,\gamma \downarrow}, & m_z = -3/2
    \end{cases} \label{Quartics}
\end{equation}
where $\alpha < \beta < \gamma$.
The basis states within the doublet ($S = 1/2$) sectors with $m_z = 1/2$ are given by 
\begin{equation}
    \ket{D_{1/2},\alpha,\beta,\gamma,\pm} = 
    \begin{cases}
    \frac{1}{\sqrt{3}}\left(
        \ket{\alpha\downarrow,\beta\uparrow,\gamma \uparrow} 
        +  e^{\pm i 2\pi/3}\ket{\alpha\uparrow,\beta\downarrow,\gamma \uparrow}
        + e^{\mp i 2\pi/3}\ket{\alpha\uparrow,\beta\uparrow,\gamma \downarrow}
        \right), & \alpha \neq \beta \neq \gamma \\
        \ket{\alpha \uparrow, \alpha \downarrow, \gamma \uparrow}, & \alpha = \beta \neq \gamma \\
        \ket{\alpha \uparrow, \beta \uparrow, \beta \downarrow}, & \alpha \neq \beta = \gamma
\end{cases} \label{Doublets}
\end{equation}
\end{widetext}
where $\alpha \leq \beta \leq \gamma$ (excluding $\alpha = \beta = \gamma$), and the $\pm$ in Eq. (\ref{Doublets}) is a chirality quantum number. 
This extra quantum number is a consequence of there being two $S = 1/2$ sectors when combining $N = 3$ electrons. 
The $S = 1/2$ basis states with $m_z = -1/2$ can be found by applying the $S_-$ operator to the states given in Eq. (\ref{Doublets}).

\section{Triangular plaquette effective Hamiltonian}\label{TriApp}

In Eq. (\ref{HST}) the main text, we provided the singlet and triplet blocks for the low-energy Hamiltonian of the triangular plaquette shown in Fig. \ref{FIGresults1}(a) with  $M=2$ electron present.
Here, we provide details leading to that Hamiltonian.

The starting point is the effective Hamiltonian $H_{eff}^{\T{g.v.}}$ given in Eq. (\ref{Hgv}) of the main text. This effective Hamiltonian contains a projection operator $P$ that excludes states double occupancy of any dots and excited valley occupation. 
With this fact, along with Eq. (\ref{Singlet}),
we can write down the allowed spin singlet ($S = 0$) states. 
These are
\begin{align}
    \ket{S,1_g,2_g} = \frac{1}{\sqrt{2}}\left(
    \ket{1_g \uparrow, 2_g \downarrow}
    - \ket{1_g \downarrow, 2_g \uparrow}
    \right),  \label{S1g2g}\\
    \ket{S,1_g,3_g} = \frac{1}{\sqrt{2}}\left(
    \ket{1_g \uparrow, 3_g \downarrow}
    - \ket{1_g \downarrow, 3_g \uparrow}
    \right),  \label{S1g3g}\\
    \ket{S,2_g,3_g} = \frac{1}{\sqrt{2}}\left(
    \ket{2_g \uparrow, 3_g \downarrow}
    - \ket{2_g \downarrow, 3_g \uparrow}
    \right), \label{S2g3g}
\end{align}
where $i_g$ indicates the ground valley of dot $i$.
Direct calculation of the matrix elements of $H_{eff}^{\T{g.v.}}$ in the basis of Eqs. (\ref{S1g2g}~-~ \ref{S2g3g}) then yields
\begin{equation}
    H_S^{\triangle} = \begin{pmatrix}
        \varepsilon_{1}^\prime + \varepsilon_{2}^\prime &  t_{3,2}^\prime & t_{1,3}^\prime \\
         \ t_{3,2}^\prime & \varepsilon_{1}^\prime + \varepsilon_{3}^\prime & t_{2,1}^\prime \\
        t_{1,3}^\prime & 
        t_{2,1}^\prime &
        \varepsilon_{2}^\prime + \varepsilon_{3}^\prime 
    \end{pmatrix}.
\end{equation}
Using Eq. (\ref{Triplets}), we can write next write down the allowed triplet ($S = 1$) states by the projection operator $P$. For the $m_z = 1$ sector, these are
\begin{align}
    \ket{T_1,1_g,2_g} = \ket{1_g \uparrow, 2_g \uparrow}, \label{T1g2g}\\
    \ket{T_1,1_g,3_g} = \ket{1_g \uparrow, 3_g \uparrow}, \\
    \ket{T_1,2_g,3_g} = \ket{2_g \uparrow, 3_g \uparrow}. \label{T2g3g}
\end{align}
Direct calculation of the matrix elements of $H_{eff}^{\T{g.v.}}$ in the basis of Eqs. (\ref{T1g2g}~-~\ref{T2g3g}) then yields
\begin{equation}
    H_T^{\triangle} = \begin{pmatrix}
        \varepsilon_{1}^\prime + \varepsilon_{2}^\prime &  t_{3,2}^\prime & -t_{1,3}^\prime \\
         \ t_{3,2}^\prime & \varepsilon_{1}^\prime + \varepsilon_{3}^\prime & t_{2,1}^\prime \\
        -t_{1,3}^\prime & 
        t_{2,1}^\prime &
        \varepsilon_{2}^\prime + \varepsilon_{3}^\prime 
    \end{pmatrix}, \label{HTTri}
\end{equation}
where the negative sign in the $t_{1,3}^\prime$ element is due to the anti-commutation of fermionic creation/annihilation operators.
Note that the triplet Hamiltonian blocks with $m_z = -1,0$ take the same form as Eq. (\ref{HTTri}) due to the $SU(2)$ symmetry.
For convenience, we transfer the negative sign from the $t_{1,3}^\prime$ elements to the $t_{3,2}^\prime$ elements in Eq. (\ref{HTTri}) by multiplying the third triplet basis state in Eq. (\ref{T2g3g}) by $-1$.
With this final step, we arrive at the final form of the singlet and triplet Hamiltonian blocks given in Eq. (\ref{HST}) of the main text.

\section{Broken Nagaoka positivity condition in square plaquette} \label{BNPCApp}

In Sec. \ref{SqPlaq} of the main text, we stated that the Nagaoka positivity condition (NPC) for a square plaquette is broken in $1/3$ of valley phase parameter space. 
In this appendix we derive this result.

To assess whether the NPC is broken, we need to assess the sign of the ground-valley hoppings $t_{i,j}^\prime = t_{i,j}\cos(\phi_{i,j}/2)$ in the low-energy theory in Eq. (\ref{Hgv}) of the main text.
Without loss of generality, we assume that $\phi_1 = 0$. Therefore, $t_{2,1}^\prime, t^\prime_{1,4} < 0$, and the NPC will be broken if either $t_{3,2}^\prime > 0$ or $t_{4,3}^\prime > 0$, but not both.
For a given $\phi_4$, we can map out the region of the $(\phi_2,\phi_3)$-space in which the NPC is broken.
For example, the NPC is broken for the case of $\phi_4 = \pi/2$ in the blue regions of Fig. \ref{FIGresults1}(d) of the main text.
Generically, the height of the bottom blue region is given by $\phi_4$ if $\phi_4 > 0$.
From these considerations, we deduce that the fraction of the area with a broken NPC is given by
\begin{equation}
    f(\phi_4) = \frac{1}{4} 
    \left(
    1 + \frac{\phi_4^2}{\pi^2}
    \right), \label{fphi4}
\end{equation}
for any given value of $\phi_{4}$.
The total fraction of the entire valley parameter space in which the NPC is broken is then found by averaging Eq. (\ref{fphi4}) over all possible values of $\phi_4$. Namely,
\begin{equation}
    \begin{split}
    F =& \frac{1}{2\pi} \int_{-\pi}^\pi f(\phi_4) \, d\phi_4, \\
    =& 1/3,
    \end{split}
\end{equation}
which is the fraction of valley phase parameter space stated in Sec. \ref{SqPlaq} of the main text.

\section{Square plaquette effective Hamiltonian for $M = 2$ electrons}\label{SqApp}
In Eq. (\ref{HSTSq}) of the main text, we provided the singlet and triplet blocks for the low-energy Hamiltonian of $M = 2$ electrons in the square plaquette shown in Fig. \ref{FIGresults1}(c). Here, we provide details leading to that Hamiltonian along with full expressions for its $A, B,$ and $C_\pm$ parameters.

In contrast to the triangular plaquette, extended Coulomb interactions play an important role in the square plaquette with $M = 2$ electrons present. 
Indeed, under the assumption that $V \gg |t_{i,j}|, |\Delta_i|$, minimization of the Coulomb energy is what determines the low-energy subspace.  
Specifically, $\{
\plaquette{white}{black}{black}{white},
\plaquette{black}{white}{white}{black}  
\}$ are the low-energy charge configurations, while 
$\{ 
\plaquette{white}{white}{black}{black},
\plaquette{black}{white}{black}{white},
\plaquette{white}{black}{white}{black},
\plaquette{black}{black}{white}{white}
\}$ are high-energy charge configurations.
As in the main text, a black dot indicates the presence of an electron.
If the valley splittings of each dot are large compared to the potential energy difference between the $2$ low-energy charge configurations, $(\varepsilon_{1}^\prime + \varepsilon_{3}^\prime) - (\varepsilon_{2}^\prime + \varepsilon_{4}^\prime)$, 
the relevant low-energy subspace contains states in the low-energy charge configurations with exclusively  ground valleys occupied.
For the singlet ($S = 0$) and triplet ($S = 1$) sector, these are given by $\{\ket{S,1_g,3_g}, \ket{S,2_g, 4_g}\}$ and $\{\ket{T_{m_z},1_g,3_g}, \ket{T_{m_z},2_g, 4_g}\}$, respectively, where the notation of Eqs. (\ref{Triplets}, \ref{Singlet}) is being used.
To obtain an effective Hamiltonian for these low-energy subspaces, we perform a second-order Schrieffer-Wolff transformation, which involves a summation over second-order perturbation pathways involving the high-energy charge configurations as virtual states.
Note that these second-order pathways must also involve states with excited valley occupied
After all, the additional energy to excite an electron from a ground valley to excited valley is assumed small compared to the nearest-neighbor Coulomb interaction energy $V$.
However, we find that second-order pathways that provide off-diagonal coupling between the two low-energy states in either the singlet or triplet sector can only involve hoppings between ground valleys of neighboring dots. 
Summing over all second-order pathways, we find an effective Hamiltonian whose form is given in Eq. (\ref{HSTSq}) of the main text. The $A$ parameter is given by
\begin{widetext}
\begin{equation}
\begin{split}
    A =
    &-\frac{(t_{3,2}^\prime)^2}{V + \varepsilon_2^\prime - \varepsilon_{3}^\prime}
    -\frac{(t_{4,3}^\prime)^2}{V + \varepsilon_4^\prime - \varepsilon_{3}^\prime}
    -\frac{(t_{2,1}^\prime)^2}{V + \varepsilon_2^\prime - \varepsilon_{1}^\prime}
    -\frac{(t_{1,4}^\prime)^2}{V + \varepsilon_4^\prime - \varepsilon_{1}^\prime} \\
    &-\frac{(t_{3,2} \sin(\phi_{3,2}/2))^2}{V + \varepsilon_2^\prime - \varepsilon_{3}^\prime + 2|\Delta_2|}
    -\frac{(t_{4,3} \sin(\phi_{4,3}/2))^2}{V + \varepsilon_4^\prime - \varepsilon_{3}^\prime+ 2|\Delta_4|}
    -\frac{(t_{2,1} \sin(\phi_{2,1}/2))^2}{V + \varepsilon_2^\prime - \varepsilon_{1}^\prime + 2|\Delta_2|}
    -\frac{(t_{1,4} \sin(\phi_{1,4}/2))^2}{V + \varepsilon_4^\prime - \varepsilon_{1}^\prime+2|\Delta_4|},  
    \end{split}
\end{equation}
where the first line involves hopping to excited states with only ground valleys occupied, while the second lines involves excited states with an excited valley occupied. The $B$ parameter is given by
\begin{equation}
\begin{split}
    B =
    &-\frac{(t_{1,4}^\prime)^2}{V + \varepsilon_1^\prime - \varepsilon_{4}^\prime}
    -\frac{(t_{2,1}^\prime)^2}{V + \varepsilon_1^\prime - \varepsilon_{2}^\prime}
    -\frac{(t_{4,3}^\prime)^2}{V + \varepsilon_3^\prime - \varepsilon_{4}^\prime}
    -\frac{(t_{3,2}^\prime)^2}{V + \varepsilon_3^\prime - \varepsilon_{2}^\prime} \\
    &-\frac{(t_{1,4} \sin(\phi_{1,4}/2))^2}{V + \varepsilon_1^\prime - \varepsilon_{4}^\prime + 2|\Delta_1|}
    -\frac{(t_{2,1} \sin(\phi_{2,1}/2))^2}{V + \varepsilon_1^\prime - \varepsilon_{2}^\prime+ 2|\Delta_1|}
    -\frac{(t_{4,3} \sin(\phi_{4,3}/2))^2}{V + \varepsilon_3^\prime - \varepsilon_{4}^\prime + 2|\Delta_3|}
    -\frac{(t_{3,2} \sin(\phi_{1,4}/2))^2}{V + \varepsilon_3^\prime - \varepsilon_{2}^\prime+2|\Delta_3|},  
    \end{split}
\end{equation}
and $C_\pm$ is given by
\begin{equation}
    \begin{split}
    C_\pm = 
    &\mp\frac{t_{3,2}^\prime t_{1,4}^\prime}{2}
    \left(
    \frac{1}{V+\varepsilon_{2}^\prime - \varepsilon_{3}^\prime}
    +\frac{1}{V+\varepsilon_{1}^\prime - \varepsilon_{4}^\prime}
    \right)
    -\frac{t_{4,3}^\prime t_{2,1}^\prime}{2}
    \left(
    \frac{1}{V+\varepsilon_{4}^\prime - \varepsilon_{3}^\prime}
    +\frac{1}{V+\varepsilon_{1}^\prime - \varepsilon_{2}^\prime}
    \right) \\
    &-\frac{t_{2,1}^\prime t_{4,3}^\prime}{2}
    \left(
    \frac{1}{V+\varepsilon_{2}^\prime - \varepsilon_{1}^\prime}
    +\frac{1}{V+\varepsilon_{3}^\prime - \varepsilon_{4}^\prime}
    \right)
    \mp\frac{t_{1,4}^\prime t_{3,2}^\prime}{2}
    \left(
    \frac{1}{V+\varepsilon_{4}^\prime - \varepsilon_{1}^\prime}
    +\frac{1}{V+\varepsilon_{3}^\prime - \varepsilon_{2}^\prime}
    \right).
    \end{split}
\end{equation}
\end{widetext}
For the simplified case of an unbiased plaquette, where $\varepsilon_{i}^\prime = 0$ for $i = 1,2,3,4$, we have
\begin{align}
    A = B &= -\sum_{i,j} \frac{t_{i,j}^2}{2V}, \label{ABForm} \\ 
    C_\pm &= \frac{2}{V}\left(t^{\prime}_{2,1}t^{\prime}_{4,3} \pm t^{\prime}_{3,2}t^{\prime}_{4,3}\right), \label{CForm}
\end{align}
where we have ignored contributions in A and B that are $\mathcal{O}(V^{-2})$. Eqs. (\ref{ABForm}, \ref{CForm}) are the final expressions for $A$ and $B$, and $C_\pm$ given directly above Eq. (\ref{Cmain}) and in Eq. (\ref{Cmain}), respectively, of the main text.

\section{Dependence of singlet-triplet splitting $E_{ST}$ on valley splitting} \label{ESTValleyApp}
In Eq. (\ref{EST}) of the main text, we provided the singlet-triplet splitting $E_{ST}$ in the regime where the detuning $\varepsilon_{3}^\prime > 0$ of dot 3 is comparable to the valley splitting $2|\Delta_3|$ of dot 3. In this appendix, we provide details regarding the perturbation calculation leading to this result.

In the case of $\varepsilon_{1}^\prime = \varepsilon_{2}^\prime = 0$ and $\varepsilon_{3}^\prime \gtrsim |t_{i,j}|$, the relative low-energy $M = 2$ electron states with $m_z = 0$ for the triangular plaquette shown in Fig. \ref{FIGresults1}(a) are $\{\ket{1_g \uparrow,2_g \downarrow}, \ket{1_g \downarrow,2_g \uparrow}\}$.
High-energy states include states with double occupancy, those where an excited valley in dots 1 or 2 is occupied, and those where either the ground or excited valley of dot 3 is occupied.
An example second-order perturbation pathways connecting the low-energy states have the form $\ket{1_g \uparrow, 2_g \downarrow} \rightarrow \ket{1_g \uparrow,1_g \downarrow} \rightarrow \ket{1_g \downarrow, 2_g \uparrow}$, where the intermediate state has an energy of $U$ higher than the low-energy states. 
Such pathways represent the ``conventional'' exchange interaction mechanism between two dots \cite{Burkard1999,Burkard2023}. Summing such pathways together yields a direct exchange given by
\begin{equation}
    J_{\T{direct}} = \frac{4(t_{2,1}^\prime)^{2}}{U}. \label{Jdirect}
\end{equation}
Third-order processes involving states with an electron in dot $3$ also need to be accounted for and can contribute a negative exchange for certain values of the relative valley phases between the three dots.
Example third-order pathways are shown in Fig. \ref{FIGPathways}(a) of the main text. 
Importantly, states with the ground valley of dot 3 occupied (such as $\ket{2_g\downarrow,3_g\uparrow}$) and the excited valley of dot 3 occupied (such as $\ket{2_g\downarrow,3_e\uparrow}$) need to be included in the calculation. 
These states have a unperturbed energy of $\varepsilon_3^\prime$ and $\varepsilon_3^\prime + 2|\Delta_3|$, respectively.
Summing over these third-order processes and adding them to Eq. (\ref{Jdirect}) leads to the singlet-triplet splitting given in Eq. (\ref{EST}) of the main text.



\begin{thebibliography}{67}%
\makeatletter
\providecommand \@ifxundefined [1]{%
 \@ifx{#1\undefined}
}%
\providecommand \@ifnum [1]{%
 \ifnum #1\expandafter \@firstoftwo
 \else \expandafter \@secondoftwo
 \fi
}%
\providecommand \@ifx [1]{%
 \ifx #1\expandafter \@firstoftwo
 \else \expandafter \@secondoftwo
 \fi
}%
\providecommand \natexlab [1]{#1}%
\providecommand \enquote  [1]{``#1''}%
\providecommand \bibnamefont  [1]{#1}%
\providecommand \bibfnamefont [1]{#1}%
\providecommand \citenamefont [1]{#1}%
\providecommand \href@noop [0]{\@secondoftwo}%
\providecommand \href [0]{\begingroup \@sanitize@url \@href}%
\providecommand \@href[1]{\@@startlink{#1}\@@href}%
\providecommand \@@href[1]{\endgroup#1\@@endlink}%
\providecommand \@sanitize@url [0]{\catcode `\\12\catcode `\$12\catcode `\&12\catcode `\#12\catcode `\^12\catcode `\_12\catcode `\%12\relax}%
\providecommand \@@startlink[1]{}%
\providecommand \@@endlink[0]{}%
\providecommand \url  [0]{\begingroup\@sanitize@url \@url }%
\providecommand \@url [1]{\endgroup\@href {#1}{\urlprefix }}%
\providecommand \urlprefix  [0]{URL }%
\providecommand \Eprint [0]{\href }%
\providecommand \doibase [0]{https://doi.org/}%
\providecommand \selectlanguage [0]{\@gobble}%
\providecommand \bibinfo  [0]{\@secondoftwo}%
\providecommand \bibfield  [0]{\@secondoftwo}%
\providecommand \translation [1]{[#1]}%
\providecommand \BibitemOpen [0]{}%
\providecommand \bibitemStop [0]{}%
\providecommand \bibitemNoStop [0]{.\EOS\space}%
\providecommand \EOS [0]{\spacefactor3000\relax}%
\providecommand \BibitemShut  [1]{\csname bibitem#1\endcsname}%
\let\auto@bib@innerbib\@empty
\bibitem [{\citenamefont {Loss}\ and\ \citenamefont {DiVincenzo}(1998)}]{Loss1998}%
  \BibitemOpen
  \bibfield  {author} {\bibinfo {author} {\bibfnamefont {D.}~\bibnamefont {Loss}}\ and\ \bibinfo {author} {\bibfnamefont {D.~P.}\ \bibnamefont {DiVincenzo}},\ }\bibfield  {title} {\bibinfo {title} {Quantum computation with quantum dots},\ }\href {https://doi.org/10.1103/PhysRevA.57.120} {\bibfield  {journal} {\bibinfo  {journal} {Phys. Rev. A}\ }\textbf {\bibinfo {volume} {57}},\ \bibinfo {pages} {120} (\bibinfo {year} {1998})}\BibitemShut {NoStop}%
\bibitem [{\citenamefont {Kloeffel}\ and\ \citenamefont {Loss}(2013)}]{Kloeffel2013}%
  \BibitemOpen
  \bibfield  {author} {\bibinfo {author} {\bibfnamefont {C.}~\bibnamefont {Kloeffel}}\ and\ \bibinfo {author} {\bibfnamefont {D.}~\bibnamefont {Loss}},\ }\bibfield  {title} {\bibinfo {title} {Prospects for spin-based quantum computing in quantum dots},\ }\href {https://doi.org/10.1146/annurev-conmatphys-030212-184248} {\bibfield  {journal} {\bibinfo  {journal} {Annual Review of Condensed Matter Physics}\ }\textbf {\bibinfo {volume} {4}},\ \bibinfo {pages} {51} (\bibinfo {year} {2013})}\BibitemShut {NoStop}%
\bibitem [{\citenamefont {Chatterjee}\ \emph {et~al.}(2021)\citenamefont {Chatterjee}, \citenamefont {Stevenson}, \citenamefont {Franceschi}, \citenamefont {Morello}, \citenamefont {de~Leon},\ and\ \citenamefont {Kuemmeth}}]{Chatterjee2021}%
  \BibitemOpen
  \bibfield  {author} {\bibinfo {author} {\bibfnamefont {A.}~\bibnamefont {Chatterjee}}, \bibinfo {author} {\bibfnamefont {P.}~\bibnamefont {Stevenson}}, \bibinfo {author} {\bibfnamefont {S.~D.}\ \bibnamefont {Franceschi}}, \bibinfo {author} {\bibfnamefont {A.}~\bibnamefont {Morello}}, \bibinfo {author} {\bibfnamefont {N.~P.}\ \bibnamefont {de~Leon}},\ and\ \bibinfo {author} {\bibfnamefont {F.}~\bibnamefont {Kuemmeth}},\ }\bibfield  {title} {\bibinfo {title} {Semiconductor qubits in practice},\ }\href {https://doi.org/10.1038/s42254-021-00283-9} {\bibfield  {journal} {\bibinfo  {journal} {Nature Reviews Physics}\ }\textbf {\bibinfo {volume} {3}},\ \bibinfo {pages} {157} (\bibinfo {year} {2021})}\BibitemShut {NoStop}%
\bibitem [{\citenamefont {Stano}\ and\ \citenamefont {Loss}(2022)}]{Stano2022}%
  \BibitemOpen
  \bibfield  {author} {\bibinfo {author} {\bibfnamefont {P.}~\bibnamefont {Stano}}\ and\ \bibinfo {author} {\bibfnamefont {D.}~\bibnamefont {Loss}},\ }\bibfield  {title} {\bibinfo {title} {Review of performance metrics of spin qubits in gated semiconducting nanostructures},\ }\href@noop {} {\bibfield  {journal} {\bibinfo  {journal} {Nature Reviews Physics}\ }\textbf {\bibinfo {volume} {4}},\ \bibinfo {pages} {672} (\bibinfo {year} {2022})}\BibitemShut {NoStop}%
\bibitem [{\citenamefont {Barthelemy}\ and\ \citenamefont {Vandersypen}(2013)}]{Barthelemy2013}%
  \BibitemOpen
  \bibfield  {author} {\bibinfo {author} {\bibfnamefont {P.}~\bibnamefont {Barthelemy}}\ and\ \bibinfo {author} {\bibfnamefont {L.~M.~K.}\ \bibnamefont {Vandersypen}},\ }\bibfield  {title} {\bibinfo {title} {Quantum dot systems: a versatile platform for quantum simulations},\ }\href {https://doi.org/10.1002/andp.201300124} {\bibfield  {journal} {\bibinfo  {journal} {Annalen der Physik}\ }\textbf {\bibinfo {volume} {525}},\ \bibinfo {pages} {808} (\bibinfo {year} {2013})}\BibitemShut {NoStop}%
\bibitem [{\citenamefont {Hensgens}\ \emph {et~al.}(2017)\citenamefont {Hensgens}, \citenamefont {Fujita}, \citenamefont {Janssen}, \citenamefont {Li}, \citenamefont {Diepen}, \citenamefont {Reichl}, \citenamefont {Wegscheider}, \citenamefont {Sarma},\ and\ \citenamefont {Vandersypen}}]{Hensgens2017}%
  \BibitemOpen
  \bibfield  {author} {\bibinfo {author} {\bibfnamefont {T.}~\bibnamefont {Hensgens}}, \bibinfo {author} {\bibfnamefont {T.}~\bibnamefont {Fujita}}, \bibinfo {author} {\bibfnamefont {L.}~\bibnamefont {Janssen}}, \bibinfo {author} {\bibfnamefont {X.}~\bibnamefont {Li}}, \bibinfo {author} {\bibfnamefont {C.~J.~V.}\ \bibnamefont {Diepen}}, \bibinfo {author} {\bibfnamefont {C.}~\bibnamefont {Reichl}}, \bibinfo {author} {\bibfnamefont {W.}~\bibnamefont {Wegscheider}}, \bibinfo {author} {\bibfnamefont {S.~D.}\ \bibnamefont {Sarma}},\ and\ \bibinfo {author} {\bibfnamefont {L.~M.~K.}\ \bibnamefont {Vandersypen}},\ }\bibfield  {title} {\bibinfo {title} {Quantum simulation of a {Fermi–Hubbard} model using a semiconductor quantum dot array},\ }\href {https://doi.org/10.1038/nature23022} {\bibfield  {journal} {\bibinfo  {journal} {Nature}\ }\textbf {\bibinfo {volume} {548}},\ \bibinfo {pages} {70} (\bibinfo {year} {2017})}\BibitemShut {NoStop}%
\bibitem [{\citenamefont {Kim}\ \emph {et~al.}(2022)\citenamefont {Kim}, \citenamefont {Nichol}, \citenamefont {Jordan},\ and\ \citenamefont {Franco}}]{Kim2022}%
  \BibitemOpen
  \bibfield  {author} {\bibinfo {author} {\bibfnamefont {C.~W.}\ \bibnamefont {Kim}}, \bibinfo {author} {\bibfnamefont {J.~M.}\ \bibnamefont {Nichol}}, \bibinfo {author} {\bibfnamefont {A.~N.}\ \bibnamefont {Jordan}},\ and\ \bibinfo {author} {\bibfnamefont {I.}~\bibnamefont {Franco}},\ }\bibfield  {title} {\bibinfo {title} {Analog quantum simulation of the dynamics of open quantum systems with quantum dots and microelectronic circuits},\ }\href {https://doi.org/10.1103/PRXQuantum.3.040308} {\bibfield  {journal} {\bibinfo  {journal} {PRX Quantum}\ }\textbf {\bibinfo {volume} {3}},\ \bibinfo {pages} {040308} (\bibinfo {year} {2022})}\BibitemShut {NoStop}%
\bibitem [{\citenamefont {Wang}\ \emph {et~al.}(2023)\citenamefont {Wang}, \citenamefont {Déprez}, \citenamefont {Tidjani}, \citenamefont {Lawrie}, \citenamefont {Hendrickx}, \citenamefont {Sammak}, \citenamefont {Scappucci},\ and\ \citenamefont {Veldhorst}}]{Wang2023}%
  \BibitemOpen
  \bibfield  {author} {\bibinfo {author} {\bibfnamefont {C.-A.}\ \bibnamefont {Wang}}, \bibinfo {author} {\bibfnamefont {C.}~\bibnamefont {Déprez}}, \bibinfo {author} {\bibfnamefont {H.}~\bibnamefont {Tidjani}}, \bibinfo {author} {\bibfnamefont {W.~I.~L.}\ \bibnamefont {Lawrie}}, \bibinfo {author} {\bibfnamefont {N.~W.}\ \bibnamefont {Hendrickx}}, \bibinfo {author} {\bibfnamefont {A.}~\bibnamefont {Sammak}}, \bibinfo {author} {\bibfnamefont {G.}~\bibnamefont {Scappucci}},\ and\ \bibinfo {author} {\bibfnamefont {M.}~\bibnamefont {Veldhorst}},\ }\bibfield  {title} {\bibinfo {title} {Probing resonating valence bonds on a programmable germanium quantum simulator},\ }\href {https://doi.org/10.1038/s41534-023-00727-3} {\bibfield  {journal} {\bibinfo  {journal} {npj Quantum Information}\ }\textbf {\bibinfo {volume} {9}},\ \bibinfo {pages} {58} (\bibinfo {year} {2023})}\BibitemShut {NoStop}%
\bibitem [{\citenamefont {Mills}\ \emph {et~al.}(2022)\citenamefont {Mills}, \citenamefont {Guinn}, \citenamefont {Gullans}, \citenamefont {Sigillito}, \citenamefont {Feldman}, \citenamefont {Nielsen},\ and\ \citenamefont {Petta}}]{Mills2021}%
  \BibitemOpen
  \bibfield  {author} {\bibinfo {author} {\bibfnamefont {A.~R.}\ \bibnamefont {Mills}}, \bibinfo {author} {\bibfnamefont {C.~R.}\ \bibnamefont {Guinn}}, \bibinfo {author} {\bibfnamefont {M.~J.}\ \bibnamefont {Gullans}}, \bibinfo {author} {\bibfnamefont {A.~J.}\ \bibnamefont {Sigillito}}, \bibinfo {author} {\bibfnamefont {M.~M.}\ \bibnamefont {Feldman}}, \bibinfo {author} {\bibfnamefont {E.}~\bibnamefont {Nielsen}},\ and\ \bibinfo {author} {\bibfnamefont {J.~R.}\ \bibnamefont {Petta}},\ }\bibfield  {title} {\bibinfo {title} {Two-qubit silicon quantum processor with operation fidelity exceeding 99\%},\ }\href {https://doi.org/10.1126/sciadv.abn5130} {\bibfield  {journal} {\bibinfo  {journal} {Science Advances}\ }\textbf {\bibinfo {volume} {8}},\ \bibinfo {pages} {eabn5130} (\bibinfo {year} {2022})}\BibitemShut {NoStop}%
\bibitem [{\citenamefont {Xue}\ \emph {et~al.}(2022{\natexlab{a}})\citenamefont {Xue}, \citenamefont {Russ}, \citenamefont {Samkharadze}, \citenamefont {Undseth}, \citenamefont {Sammak}, \citenamefont {Scappucci},\ and\ \citenamefont {Vandersypen}}]{Xue2022}%
  \BibitemOpen
  \bibfield  {author} {\bibinfo {author} {\bibfnamefont {X.}~\bibnamefont {Xue}}, \bibinfo {author} {\bibfnamefont {M.}~\bibnamefont {Russ}}, \bibinfo {author} {\bibfnamefont {N.}~\bibnamefont {Samkharadze}}, \bibinfo {author} {\bibfnamefont {B.}~\bibnamefont {Undseth}}, \bibinfo {author} {\bibfnamefont {A.}~\bibnamefont {Sammak}}, \bibinfo {author} {\bibfnamefont {G.}~\bibnamefont {Scappucci}},\ and\ \bibinfo {author} {\bibfnamefont {L.~M.~K.}\ \bibnamefont {Vandersypen}},\ }\bibfield  {title} {\bibinfo {title} {Quantum logic with spin qubits crossing the surface code threshold},\ }\href {https://doi.org/10.1038/s41586-021-04273-w} {\bibfield  {journal} {\bibinfo  {journal} {Nature}\ }\textbf {\bibinfo {volume} {601}},\ \bibinfo {pages} {343} (\bibinfo {year} {2022}{\natexlab{a}})}\BibitemShut {NoStop}%
\bibitem [{\citenamefont {Noiri}\ \emph {et~al.}(2022)\citenamefont {Noiri}, \citenamefont {Takeda}, \citenamefont {Nakajima}, \citenamefont {Kobayashi}, \citenamefont {Sammak}, \citenamefont {Scappucci},\ and\ \citenamefont {Tarucha}}]{Noiri2022}%
  \BibitemOpen
  \bibfield  {author} {\bibinfo {author} {\bibfnamefont {A.}~\bibnamefont {Noiri}}, \bibinfo {author} {\bibfnamefont {K.}~\bibnamefont {Takeda}}, \bibinfo {author} {\bibfnamefont {T.}~\bibnamefont {Nakajima}}, \bibinfo {author} {\bibfnamefont {T.}~\bibnamefont {Kobayashi}}, \bibinfo {author} {\bibfnamefont {A.}~\bibnamefont {Sammak}}, \bibinfo {author} {\bibfnamefont {G.}~\bibnamefont {Scappucci}},\ and\ \bibinfo {author} {\bibfnamefont {S.}~\bibnamefont {Tarucha}},\ }\bibfield  {title} {\bibinfo {title} {Fast universal quantum gate above the fault-tolerance threshold in silicon},\ }\href {https://doi.org/10.1038/s41586-021-04182-y} {\bibfield  {journal} {\bibinfo  {journal} {Nature}\ }\textbf {\bibinfo {volume} {601}},\ \bibinfo {pages} {338} (\bibinfo {year} {2022})}\BibitemShut {NoStop}%
\bibitem [{\citenamefont {Fowler}\ \emph {et~al.}(2012)\citenamefont {Fowler}, \citenamefont {Mariantoni}, \citenamefont {Martinis},\ and\ \citenamefont {Cleland}}]{Fowler2012}%
  \BibitemOpen
  \bibfield  {author} {\bibinfo {author} {\bibfnamefont {A.~G.}\ \bibnamefont {Fowler}}, \bibinfo {author} {\bibfnamefont {M.}~\bibnamefont {Mariantoni}}, \bibinfo {author} {\bibfnamefont {J.~M.}\ \bibnamefont {Martinis}},\ and\ \bibinfo {author} {\bibfnamefont {A.~N.}\ \bibnamefont {Cleland}},\ }\bibfield  {title} {\bibinfo {title} {Surface codes: Towards practical large-scale quantum computation},\ }\href {https://doi.org/10.1103/PhysRevA.86.032324} {\bibfield  {journal} {\bibinfo  {journal} {Phys. Rev. A}\ }\textbf {\bibinfo {volume} {86}},\ \bibinfo {pages} {032324} (\bibinfo {year} {2012})}\BibitemShut {NoStop}%
\bibitem [{\citenamefont {Burkard}\ \emph {et~al.}(2023)\citenamefont {Burkard}, \citenamefont {Ladd}, \citenamefont {Pan}, \citenamefont {Nichol},\ and\ \citenamefont {Petta}}]{Burkard2023}%
  \BibitemOpen
  \bibfield  {author} {\bibinfo {author} {\bibfnamefont {G.}~\bibnamefont {Burkard}}, \bibinfo {author} {\bibfnamefont {T.~D.}\ \bibnamefont {Ladd}}, \bibinfo {author} {\bibfnamefont {A.}~\bibnamefont {Pan}}, \bibinfo {author} {\bibfnamefont {J.~M.}\ \bibnamefont {Nichol}},\ and\ \bibinfo {author} {\bibfnamefont {J.~R.}\ \bibnamefont {Petta}},\ }\bibfield  {title} {\bibinfo {title} {Semiconductor spin qubits},\ }\href {https://doi.org/10.1103/RevModPhys.95.025003} {\bibfield  {journal} {\bibinfo  {journal} {Rev. Mod. Phys.}\ }\textbf {\bibinfo {volume} {95}},\ \bibinfo {pages} {025003} (\bibinfo {year} {2023})}\BibitemShut {NoStop}%
\bibitem [{\citenamefont {Burkard}\ \emph {et~al.}(1999)\citenamefont {Burkard}, \citenamefont {Loss},\ and\ \citenamefont {DiVincenzo}}]{Burkard1999}%
  \BibitemOpen
  \bibfield  {author} {\bibinfo {author} {\bibfnamefont {G.}~\bibnamefont {Burkard}}, \bibinfo {author} {\bibfnamefont {D.}~\bibnamefont {Loss}},\ and\ \bibinfo {author} {\bibfnamefont {D.~P.}\ \bibnamefont {DiVincenzo}},\ }\bibfield  {title} {\bibinfo {title} {Coupled quantum dots as quantum gates},\ }\href {https://doi.org/10.1103/PhysRevB.59.2070} {\bibfield  {journal} {\bibinfo  {journal} {Phys. Rev. B}\ }\textbf {\bibinfo {volume} {59}},\ \bibinfo {pages} {2070} (\bibinfo {year} {1999})}\BibitemShut {NoStop}%
\bibitem [{\citenamefont {Petta}\ \emph {et~al.}(2005)\citenamefont {Petta}, \citenamefont {Johnson}, \citenamefont {Taylor}, \citenamefont {Laird}, \citenamefont {Yacoby}, \citenamefont {Lukin}, \citenamefont {Marcus}, \citenamefont {Hanson},\ and\ \citenamefont {Gossard}}]{Petta2005}%
  \BibitemOpen
  \bibfield  {author} {\bibinfo {author} {\bibfnamefont {J.~R.}\ \bibnamefont {Petta}}, \bibinfo {author} {\bibfnamefont {A.~C.}\ \bibnamefont {Johnson}}, \bibinfo {author} {\bibfnamefont {J.~M.}\ \bibnamefont {Taylor}}, \bibinfo {author} {\bibfnamefont {E.~A.}\ \bibnamefont {Laird}}, \bibinfo {author} {\bibfnamefont {A.}~\bibnamefont {Yacoby}}, \bibinfo {author} {\bibfnamefont {M.~D.}\ \bibnamefont {Lukin}}, \bibinfo {author} {\bibfnamefont {C.~M.}\ \bibnamefont {Marcus}}, \bibinfo {author} {\bibfnamefont {M.~P.}\ \bibnamefont {Hanson}},\ and\ \bibinfo {author} {\bibfnamefont {A.~C.}\ \bibnamefont {Gossard}},\ }\bibfield  {title} {\bibinfo {title} {Coherent manipulation of coupled electron spins in semiconductor quantum dots},\ }\href {https://doi.org/10.1126/science.1116955} {\bibfield  {journal} {\bibinfo  {journal} {Science}\ }\textbf {\bibinfo {volume} {309}},\ \bibinfo {pages} {2180} (\bibinfo {year} {2005})}\BibitemShut {NoStop}%
\bibitem [{\citenamefont {Maune}\ \emph {et~al.}(2012)\citenamefont {Maune}, \citenamefont {Borselli}, \citenamefont {Huang}, \citenamefont {Ladd}, \citenamefont {Deelman}, \citenamefont {Holabird}, \citenamefont {Kiselev}, \citenamefont {Alvarado-Rodriguez}, \citenamefont {Ross}, \citenamefont {Schmitz} \emph {et~al.}}]{Maune2012}%
  \BibitemOpen
  \bibfield  {author} {\bibinfo {author} {\bibfnamefont {B.~M.}\ \bibnamefont {Maune}}, \bibinfo {author} {\bibfnamefont {M.~G.}\ \bibnamefont {Borselli}}, \bibinfo {author} {\bibfnamefont {B.}~\bibnamefont {Huang}}, \bibinfo {author} {\bibfnamefont {T.~D.}\ \bibnamefont {Ladd}}, \bibinfo {author} {\bibfnamefont {P.~W.}\ \bibnamefont {Deelman}}, \bibinfo {author} {\bibfnamefont {K.~S.}\ \bibnamefont {Holabird}}, \bibinfo {author} {\bibfnamefont {A.~A.}\ \bibnamefont {Kiselev}}, \bibinfo {author} {\bibfnamefont {I.}~\bibnamefont {Alvarado-Rodriguez}}, \bibinfo {author} {\bibfnamefont {R.~S.}\ \bibnamefont {Ross}}, \bibinfo {author} {\bibfnamefont {A.~E.}\ \bibnamefont {Schmitz}}, \emph {et~al.},\ }\bibfield  {title} {\bibinfo {title} {Coherent singlet-triplet oscillations in a silicon-based double quantum dot},\ }\href@noop {} {\bibfield  {journal} {\bibinfo  {journal} {Nature}\ }\textbf {\bibinfo {volume} {481}},\ \bibinfo {pages} {344} (\bibinfo {year} {2012})}\BibitemShut {NoStop}%
\bibitem [{\citenamefont {Levy}(2002)}]{Levy2002}%
  \BibitemOpen
  \bibfield  {author} {\bibinfo {author} {\bibfnamefont {J.}~\bibnamefont {Levy}},\ }\bibfield  {title} {\bibinfo {title} {Universal quantum computation with spin-$1/2$ pairs and heisenberg exchange},\ }\href {https://doi.org/10.1103/PhysRevLett.89.147902} {\bibfield  {journal} {\bibinfo  {journal} {Phys. Rev. Lett.}\ }\textbf {\bibinfo {volume} {89}},\ \bibinfo {pages} {147902} (\bibinfo {year} {2002})}\BibitemShut {NoStop}%
\bibitem [{\citenamefont {Weinstein}\ \emph {et~al.}(2023)\citenamefont {Weinstein}, \citenamefont {Reed}, \citenamefont {Jones}, \citenamefont {Andrews}, \citenamefont {Barnes}, \citenamefont {Blumoff}, \citenamefont {Euliss}, \citenamefont {Eng}, \citenamefont {Fong}, \citenamefont {Ha}, \citenamefont {Hulbert}, \citenamefont {Jackson}, \citenamefont {Jura}, \citenamefont {Keating}, \citenamefont {Kerckhoff}, \citenamefont {Kiselev}, \citenamefont {Matten}, \citenamefont {Sabbir}, \citenamefont {Smith}, \citenamefont {Wright}, \citenamefont {Rakher}, \citenamefont {Ladd},\ and\ \citenamefont {Borselli}}]{Weinstein2023}%
  \BibitemOpen
  \bibfield  {author} {\bibinfo {author} {\bibfnamefont {A.~J.}\ \bibnamefont {Weinstein}}, \bibinfo {author} {\bibfnamefont {M.~D.}\ \bibnamefont {Reed}}, \bibinfo {author} {\bibfnamefont {A.~M.}\ \bibnamefont {Jones}}, \bibinfo {author} {\bibfnamefont {R.~W.}\ \bibnamefont {Andrews}}, \bibinfo {author} {\bibfnamefont {D.}~\bibnamefont {Barnes}}, \bibinfo {author} {\bibfnamefont {J.~Z.}\ \bibnamefont {Blumoff}}, \bibinfo {author} {\bibfnamefont {L.~E.}\ \bibnamefont {Euliss}}, \bibinfo {author} {\bibfnamefont {K.}~\bibnamefont {Eng}}, \bibinfo {author} {\bibfnamefont {B.~H.}\ \bibnamefont {Fong}}, \bibinfo {author} {\bibfnamefont {S.~D.}\ \bibnamefont {Ha}}, \bibinfo {author} {\bibfnamefont {D.~R.}\ \bibnamefont {Hulbert}}, \bibinfo {author} {\bibfnamefont {C.~A.~C.}\ \bibnamefont {Jackson}}, \bibinfo {author} {\bibfnamefont {M.}~\bibnamefont {Jura}}, \bibinfo {author} {\bibfnamefont {T.~E.}\ \bibnamefont {Keating}}, \bibinfo {author} {\bibfnamefont {J.}~\bibnamefont {Kerckhoff}}, \bibinfo {author}
  {\bibfnamefont {A.~A.}\ \bibnamefont {Kiselev}}, \bibinfo {author} {\bibfnamefont {J.}~\bibnamefont {Matten}}, \bibinfo {author} {\bibfnamefont {G.}~\bibnamefont {Sabbir}}, \bibinfo {author} {\bibfnamefont {A.}~\bibnamefont {Smith}}, \bibinfo {author} {\bibfnamefont {J.}~\bibnamefont {Wright}}, \bibinfo {author} {\bibfnamefont {M.~T.}\ \bibnamefont {Rakher}}, \bibinfo {author} {\bibfnamefont {T.~D.}\ \bibnamefont {Ladd}},\ and\ \bibinfo {author} {\bibfnamefont {M.~G.}\ \bibnamefont {Borselli}},\ }\bibfield  {title} {\bibinfo {title} {Universal logic with encoded spin qubits in silicon},\ }\href {https://doi.org/10.1038/s41586-023-05777-3} {\bibfield  {journal} {\bibinfo  {journal} {Nature}\ }\textbf {\bibinfo {volume} {615}},\ \bibinfo {pages} {817} (\bibinfo {year} {2023})}\BibitemShut {NoStop}%
\bibitem [{\citenamefont {Lieb}\ and\ \citenamefont {Mattis}(1962)}]{Lieb1962}%
  \BibitemOpen
  \bibfield  {author} {\bibinfo {author} {\bibfnamefont {E.}~\bibnamefont {Lieb}}\ and\ \bibinfo {author} {\bibfnamefont {D.}~\bibnamefont {Mattis}},\ }\bibfield  {title} {\bibinfo {title} {Theory of ferromagnetism and the ordering of electronic energy levels},\ }\href {https://doi.org/10.1103/PhysRev.125.164} {\bibfield  {journal} {\bibinfo  {journal} {Phys. Rev.}\ }\textbf {\bibinfo {volume} {125}},\ \bibinfo {pages} {164} (\bibinfo {year} {1962})}\BibitemShut {NoStop}%
\bibitem [{\citenamefont {Khodjasteh}\ and\ \citenamefont {Viola}(2009)}]{Khodjasteh2009}%
  \BibitemOpen
  \bibfield  {author} {\bibinfo {author} {\bibfnamefont {K.}~\bibnamefont {Khodjasteh}}\ and\ \bibinfo {author} {\bibfnamefont {L.}~\bibnamefont {Viola}},\ }\bibfield  {title} {\bibinfo {title} {Dynamical quantum error correction of unitary operations with bounded controls},\ }\href {https://doi.org/10.1103/PhysRevA.80.032314} {\bibfield  {journal} {\bibinfo  {journal} {Phys. Rev. A}\ }\textbf {\bibinfo {volume} {80}},\ \bibinfo {pages} {032314} (\bibinfo {year} {2009})}\BibitemShut {NoStop}%
\bibitem [{\citenamefont {Wang}\ \emph {et~al.}(2012)\citenamefont {Wang}, \citenamefont {Bishop}, \citenamefont {Kestner}, \citenamefont {Barnes}, \citenamefont {Sun},\ and\ \citenamefont {Sarma}}]{Wang2012}%
  \BibitemOpen
  \bibfield  {author} {\bibinfo {author} {\bibfnamefont {X.}~\bibnamefont {Wang}}, \bibinfo {author} {\bibfnamefont {L.~S.}\ \bibnamefont {Bishop}}, \bibinfo {author} {\bibfnamefont {J.}~\bibnamefont {Kestner}}, \bibinfo {author} {\bibfnamefont {E.}~\bibnamefont {Barnes}}, \bibinfo {author} {\bibfnamefont {K.}~\bibnamefont {Sun}},\ and\ \bibinfo {author} {\bibfnamefont {S.~D.}\ \bibnamefont {Sarma}},\ }\bibfield  {title} {\bibinfo {title} {Composite pulses for robust universal control of singlet–triplet qubits},\ }\href {https://doi.org/10.1038/ncomms2003} {\bibfield  {journal} {\bibinfo  {journal} {Nature Communications}\ }\textbf {\bibinfo {volume} {3}},\ \bibinfo {pages} {997} (\bibinfo {year} {2012})}\BibitemShut {NoStop}%
\bibitem [{\citenamefont {Kestner}\ \emph {et~al.}(2013)\citenamefont {Kestner}, \citenamefont {Wang}, \citenamefont {Bishop}, \citenamefont {Barnes},\ and\ \citenamefont {Das~Sarma}}]{Kestner2013}%
  \BibitemOpen
  \bibfield  {author} {\bibinfo {author} {\bibfnamefont {J.~P.}\ \bibnamefont {Kestner}}, \bibinfo {author} {\bibfnamefont {X.}~\bibnamefont {Wang}}, \bibinfo {author} {\bibfnamefont {L.~S.}\ \bibnamefont {Bishop}}, \bibinfo {author} {\bibfnamefont {E.}~\bibnamefont {Barnes}},\ and\ \bibinfo {author} {\bibfnamefont {S.}~\bibnamefont {Das~Sarma}},\ }\bibfield  {title} {\bibinfo {title} {Noise-resistant control for a spin qubit array},\ }\href {https://doi.org/10.1103/PhysRevLett.110.140502} {\bibfield  {journal} {\bibinfo  {journal} {Phys. Rev. Lett.}\ }\textbf {\bibinfo {volume} {110}},\ \bibinfo {pages} {140502} (\bibinfo {year} {2013})}\BibitemShut {NoStop}%
\bibitem [{\citenamefont {Wang}\ \emph {et~al.}(2014)\citenamefont {Wang}, \citenamefont {Bishop}, \citenamefont {Barnes}, \citenamefont {Kestner},\ and\ \citenamefont {Sarma}}]{Wang2014}%
  \BibitemOpen
  \bibfield  {author} {\bibinfo {author} {\bibfnamefont {X.}~\bibnamefont {Wang}}, \bibinfo {author} {\bibfnamefont {L.~S.}\ \bibnamefont {Bishop}}, \bibinfo {author} {\bibfnamefont {E.}~\bibnamefont {Barnes}}, \bibinfo {author} {\bibfnamefont {J.~P.}\ \bibnamefont {Kestner}},\ and\ \bibinfo {author} {\bibfnamefont {S.~D.}\ \bibnamefont {Sarma}},\ }\bibfield  {title} {\bibinfo {title} {Robust quantum gates for singlet-triplet spin qubits using composite pulses},\ }\href {https://doi.org/10.1103/PhysRevA.89.022310} {\bibfield  {journal} {\bibinfo  {journal} {Phys. Rev. A}\ }\textbf {\bibinfo {volume} {89}},\ \bibinfo {pages} {022310} (\bibinfo {year} {2014})}\BibitemShut {NoStop}%
\bibitem [{\citenamefont {Lindemann}\ \emph {et~al.}(2002)\citenamefont {Lindemann}, \citenamefont {Ihn}, \citenamefont {Heinzel}, \citenamefont {Zwerger}, \citenamefont {Ensslin}, \citenamefont {Maranowski},\ and\ \citenamefont {Gossard}}]{Lindemann2002}%
  \BibitemOpen
  \bibfield  {author} {\bibinfo {author} {\bibfnamefont {S.}~\bibnamefont {Lindemann}}, \bibinfo {author} {\bibfnamefont {T.}~\bibnamefont {Ihn}}, \bibinfo {author} {\bibfnamefont {T.}~\bibnamefont {Heinzel}}, \bibinfo {author} {\bibfnamefont {W.}~\bibnamefont {Zwerger}}, \bibinfo {author} {\bibfnamefont {K.}~\bibnamefont {Ensslin}}, \bibinfo {author} {\bibfnamefont {K.}~\bibnamefont {Maranowski}},\ and\ \bibinfo {author} {\bibfnamefont {A.~C.}\ \bibnamefont {Gossard}},\ }\bibfield  {title} {\bibinfo {title} {Stability of spin states in quantum dots},\ }\href {https://doi.org/10.1103/PhysRevB.66.195314} {\bibfield  {journal} {\bibinfo  {journal} {Phys. Rev. B}\ }\textbf {\bibinfo {volume} {66}},\ \bibinfo {pages} {195314} (\bibinfo {year} {2002})}\BibitemShut {NoStop}%
\bibitem [{\citenamefont {Martins}\ \emph {et~al.}(2017)\citenamefont {Martins}, \citenamefont {Malinowski}, \citenamefont {Nissen}, \citenamefont {Fallahi}, \citenamefont {Gardner}, \citenamefont {Manfra}, \citenamefont {Marcus},\ and\ \citenamefont {Kuemmeth}}]{Martins2017}%
  \BibitemOpen
  \bibfield  {author} {\bibinfo {author} {\bibfnamefont {F.}~\bibnamefont {Martins}}, \bibinfo {author} {\bibfnamefont {F.~K.}\ \bibnamefont {Malinowski}}, \bibinfo {author} {\bibfnamefont {P.~D.}\ \bibnamefont {Nissen}}, \bibinfo {author} {\bibfnamefont {S.}~\bibnamefont {Fallahi}}, \bibinfo {author} {\bibfnamefont {G.~C.}\ \bibnamefont {Gardner}}, \bibinfo {author} {\bibfnamefont {M.~J.}\ \bibnamefont {Manfra}}, \bibinfo {author} {\bibfnamefont {C.~M.}\ \bibnamefont {Marcus}},\ and\ \bibinfo {author} {\bibfnamefont {F.}~\bibnamefont {Kuemmeth}},\ }\bibfield  {title} {\bibinfo {title} {Negative spin exchange in a multielectron quantum dot},\ }\href {https://doi.org/10.1103/PhysRevLett.119.227701} {\bibfield  {journal} {\bibinfo  {journal} {Phys. Rev. Lett.}\ }\textbf {\bibinfo {volume} {119}},\ \bibinfo {pages} {227701} (\bibinfo {year} {2017})}\BibitemShut {NoStop}%
\bibitem [{\citenamefont {Deng}\ \emph {et~al.}(2018)\citenamefont {Deng}, \citenamefont {Calderon-Vargas}, \citenamefont {Mayhall},\ and\ \citenamefont {Barnes}}]{Deng2018b}%
  \BibitemOpen
  \bibfield  {author} {\bibinfo {author} {\bibfnamefont {K.}~\bibnamefont {Deng}}, \bibinfo {author} {\bibfnamefont {F.~A.}\ \bibnamefont {Calderon-Vargas}}, \bibinfo {author} {\bibfnamefont {N.~J.}\ \bibnamefont {Mayhall}},\ and\ \bibinfo {author} {\bibfnamefont {E.}~\bibnamefont {Barnes}},\ }\bibfield  {title} {\bibinfo {title} {Negative exchange interactions in coupled few-electron quantum dots},\ }\href {https://doi.org/10.1103/PhysRevB.97.245301} {\bibfield  {journal} {\bibinfo  {journal} {Phys. Rev. B}\ }\textbf {\bibinfo {volume} {97}},\ \bibinfo {pages} {245301} (\bibinfo {year} {2018})}\BibitemShut {NoStop}%
\bibitem [{\citenamefont {Malinowski}\ \emph {et~al.}(2018)\citenamefont {Malinowski}, \citenamefont {Martins}, \citenamefont {Smith}, \citenamefont {Bartlett}, \citenamefont {Doherty}, \citenamefont {Nissen}, \citenamefont {Fallahi}, \citenamefont {Gardner}, \citenamefont {Manfra}, \citenamefont {Marcus},\ and\ \citenamefont {Kuemmeth}}]{Malinowski2018}%
  \BibitemOpen
  \bibfield  {author} {\bibinfo {author} {\bibfnamefont {F.~K.}\ \bibnamefont {Malinowski}}, \bibinfo {author} {\bibfnamefont {F.}~\bibnamefont {Martins}}, \bibinfo {author} {\bibfnamefont {T.~B.}\ \bibnamefont {Smith}}, \bibinfo {author} {\bibfnamefont {S.~D.}\ \bibnamefont {Bartlett}}, \bibinfo {author} {\bibfnamefont {A.~C.}\ \bibnamefont {Doherty}}, \bibinfo {author} {\bibfnamefont {P.~D.}\ \bibnamefont {Nissen}}, \bibinfo {author} {\bibfnamefont {S.}~\bibnamefont {Fallahi}}, \bibinfo {author} {\bibfnamefont {G.~C.}\ \bibnamefont {Gardner}}, \bibinfo {author} {\bibfnamefont {M.~J.}\ \bibnamefont {Manfra}}, \bibinfo {author} {\bibfnamefont {C.~M.}\ \bibnamefont {Marcus}},\ and\ \bibinfo {author} {\bibfnamefont {F.}~\bibnamefont {Kuemmeth}},\ }\bibfield  {title} {\bibinfo {title} {Spin of a multielectron quantum dot and its interaction with a neighboring electron},\ }\href {https://doi.org/10.1103/PhysRevX.8.011045} {\bibfield  {journal} {\bibinfo  {journal} {Phys. Rev. X}\ }\textbf {\bibinfo {volume}
  {8}},\ \bibinfo {pages} {011045} (\bibinfo {year} {2018})}\BibitemShut {NoStop}%
\bibitem [{\citenamefont {Deng}\ and\ \citenamefont {Barnes}(2020)}]{Deng2020}%
  \BibitemOpen
  \bibfield  {author} {\bibinfo {author} {\bibfnamefont {K.}~\bibnamefont {Deng}}\ and\ \bibinfo {author} {\bibfnamefont {E.}~\bibnamefont {Barnes}},\ }\bibfield  {title} {\bibinfo {title} {Interplay of exchange and superexchange in triple quantum dots},\ }\href {https://doi.org/10.1103/PhysRevB.102.035427} {\bibfield  {journal} {\bibinfo  {journal} {Phys. Rev. B}\ }\textbf {\bibinfo {volume} {102}},\ \bibinfo {pages} {035427} (\bibinfo {year} {2020})}\BibitemShut {NoStop}%
\bibitem [{\citenamefont {Wagner}\ \emph {et~al.}(1992)\citenamefont {Wagner}, \citenamefont {Merkt},\ and\ \citenamefont {Chaplik}}]{Wagner1992}%
  \BibitemOpen
  \bibfield  {author} {\bibinfo {author} {\bibfnamefont {M.}~\bibnamefont {Wagner}}, \bibinfo {author} {\bibfnamefont {U.}~\bibnamefont {Merkt}},\ and\ \bibinfo {author} {\bibfnamefont {A.~V.}\ \bibnamefont {Chaplik}},\ }\bibfield  {title} {\bibinfo {title} {Spin-singlet--spin-triplet oscillations in quantum dots},\ }\href {https://doi.org/10.1103/PhysRevB.45.1951} {\bibfield  {journal} {\bibinfo  {journal} {Phys. Rev. B}\ }\textbf {\bibinfo {volume} {45}},\ \bibinfo {pages} {1951} (\bibinfo {year} {1992})}\BibitemShut {NoStop}%
\bibitem [{\citenamefont {Zumb\"uhl}\ \emph {et~al.}(2004)\citenamefont {Zumb\"uhl}, \citenamefont {Marcus}, \citenamefont {Hanson},\ and\ \citenamefont {Gossard}}]{Zumbuhl2004}%
  \BibitemOpen
  \bibfield  {author} {\bibinfo {author} {\bibfnamefont {D.~M.}\ \bibnamefont {Zumb\"uhl}}, \bibinfo {author} {\bibfnamefont {C.~M.}\ \bibnamefont {Marcus}}, \bibinfo {author} {\bibfnamefont {M.~P.}\ \bibnamefont {Hanson}},\ and\ \bibinfo {author} {\bibfnamefont {A.~C.}\ \bibnamefont {Gossard}},\ }\bibfield  {title} {\bibinfo {title} {Cotunneling spectroscopy in few-electron quantum dots},\ }\href {https://doi.org/10.1103/PhysRevLett.93.256801} {\bibfield  {journal} {\bibinfo  {journal} {Phys. Rev. Lett.}\ }\textbf {\bibinfo {volume} {93}},\ \bibinfo {pages} {256801} (\bibinfo {year} {2004})}\BibitemShut {NoStop}%
\bibitem [{\citenamefont {Baruffa}\ \emph {et~al.}(2010)\citenamefont {Baruffa}, \citenamefont {Stano},\ and\ \citenamefont {Fabian}}]{Baruffa2010}%
  \BibitemOpen
  \bibfield  {author} {\bibinfo {author} {\bibfnamefont {F.}~\bibnamefont {Baruffa}}, \bibinfo {author} {\bibfnamefont {P.}~\bibnamefont {Stano}},\ and\ \bibinfo {author} {\bibfnamefont {J.}~\bibnamefont {Fabian}},\ }\bibfield  {title} {\bibinfo {title} {Spin-orbit coupling and anisotropic exchange in two-electron double quantum dots},\ }\href {https://doi.org/10.1103/PhysRevB.82.045311} {\bibfield  {journal} {\bibinfo  {journal} {Phys. Rev. B}\ }\textbf {\bibinfo {volume} {82}},\ \bibinfo {pages} {045311} (\bibinfo {year} {2010})}\BibitemShut {NoStop}%
\bibitem [{\citenamefont {Mehl}\ and\ \citenamefont {DiVincenzo}(2014)}]{Mehl2014}%
  \BibitemOpen
  \bibfield  {author} {\bibinfo {author} {\bibfnamefont {S.}~\bibnamefont {Mehl}}\ and\ \bibinfo {author} {\bibfnamefont {D.~P.}\ \bibnamefont {DiVincenzo}},\ }\bibfield  {title} {\bibinfo {title} {Inverted singlet-triplet qubit coded on a two-electron double quantum dot},\ }\href {https://doi.org/10.1103/PhysRevB.90.195424} {\bibfield  {journal} {\bibinfo  {journal} {Phys. Rev. B}\ }\textbf {\bibinfo {volume} {90}},\ \bibinfo {pages} {195424} (\bibinfo {year} {2014})}\BibitemShut {NoStop}%
\bibitem [{\citenamefont {Ha}\ \emph {et~al.}(2021)\citenamefont {Ha}, \citenamefont {Ha}, \citenamefont {Choi}, \citenamefont {Tang}, \citenamefont {Schmitz}, \citenamefont {Levendorf}, \citenamefont {Lee}, \citenamefont {Chappell}, \citenamefont {Adams}, \citenamefont {Hulbert} \emph {et~al.}}]{Ha2021}%
  \BibitemOpen
  \bibfield  {author} {\bibinfo {author} {\bibfnamefont {W.}~\bibnamefont {Ha}}, \bibinfo {author} {\bibfnamefont {S.~D.}\ \bibnamefont {Ha}}, \bibinfo {author} {\bibfnamefont {M.~D.}\ \bibnamefont {Choi}}, \bibinfo {author} {\bibfnamefont {Y.}~\bibnamefont {Tang}}, \bibinfo {author} {\bibfnamefont {A.~E.}\ \bibnamefont {Schmitz}}, \bibinfo {author} {\bibfnamefont {M.~P.}\ \bibnamefont {Levendorf}}, \bibinfo {author} {\bibfnamefont {K.}~\bibnamefont {Lee}}, \bibinfo {author} {\bibfnamefont {J.~M.}\ \bibnamefont {Chappell}}, \bibinfo {author} {\bibfnamefont {T.~S.}\ \bibnamefont {Adams}}, \bibinfo {author} {\bibfnamefont {D.~R.}\ \bibnamefont {Hulbert}}, \emph {et~al.},\ }\bibfield  {title} {\bibinfo {title} {A flexible design platform for {Si/SiGe} exchange-only qubits with low disorder},\ }\href {https://doi.org/10.1021/acs.nanolett.1c03026} {\bibfield  {journal} {\bibinfo  {journal} {Nano Letters}\ }\textbf {\bibinfo {volume} {22}},\ \bibinfo {pages} {1443} (\bibinfo {year} {2021})}\BibitemShut {NoStop}%
\bibitem [{\citenamefont {Unseld}\ \emph {et~al.}(2023)\citenamefont {Unseld}, \citenamefont {Meyer}, \citenamefont {Mądzik}, \citenamefont {Borsoi}, \citenamefont {de~Snoo}, \citenamefont {Amitonov}, \citenamefont {Sammak}, \citenamefont {Scappucci}, \citenamefont {Veldhorst},\ and\ \citenamefont {Vandersypen}}]{Unseld2023}%
  \BibitemOpen
  \bibfield  {author} {\bibinfo {author} {\bibfnamefont {F.~K.}\ \bibnamefont {Unseld}}, \bibinfo {author} {\bibfnamefont {M.}~\bibnamefont {Meyer}}, \bibinfo {author} {\bibfnamefont {M.~T.}\ \bibnamefont {Mądzik}}, \bibinfo {author} {\bibfnamefont {F.}~\bibnamefont {Borsoi}}, \bibinfo {author} {\bibfnamefont {S.~L.}\ \bibnamefont {de~Snoo}}, \bibinfo {author} {\bibfnamefont {S.~V.}\ \bibnamefont {Amitonov}}, \bibinfo {author} {\bibfnamefont {A.}~\bibnamefont {Sammak}}, \bibinfo {author} {\bibfnamefont {G.}~\bibnamefont {Scappucci}}, \bibinfo {author} {\bibfnamefont {M.}~\bibnamefont {Veldhorst}},\ and\ \bibinfo {author} {\bibfnamefont {L.~M.~K.}\ \bibnamefont {Vandersypen}},\ }\bibfield  {title} {\bibinfo {title} {A 2d quantum dot array in planar 28{Si/SiGe}},\ }\bibfield  {journal} {\bibinfo  {journal} {Applied Physics Letters}\ }\textbf {\bibinfo {volume} {123}},\ \href {https://doi.org/10.1063/5.0160847} {10.1063/5.0160847} (\bibinfo {year} {2023})\BibitemShut {NoStop}%
\bibitem [{\citenamefont {Acuna}\ \emph {et~al.}(2024)\citenamefont {Acuna}, \citenamefont {Broz}, \citenamefont {Shyamsundar}, \citenamefont {Mei}, \citenamefont {Feeney}, \citenamefont {Smetanka}, \citenamefont {Davis}, \citenamefont {Lee}, \citenamefont {Choi}, \citenamefont {Boyd}, \citenamefont {Suh}, \citenamefont {Ha}, \citenamefont {Jennings}, \citenamefont {Pan}, \citenamefont {Sanchez}, \citenamefont {Reed},\ and\ \citenamefont {Petta}}]{Acuna2024}%
  \BibitemOpen
  \bibfield  {author} {\bibinfo {author} {\bibfnamefont {E.}~\bibnamefont {Acuna}}, \bibinfo {author} {\bibfnamefont {J.~D.}\ \bibnamefont {Broz}}, \bibinfo {author} {\bibfnamefont {K.}~\bibnamefont {Shyamsundar}}, \bibinfo {author} {\bibfnamefont {A.~B.}\ \bibnamefont {Mei}}, \bibinfo {author} {\bibfnamefont {C.~P.}\ \bibnamefont {Feeney}}, \bibinfo {author} {\bibfnamefont {V.}~\bibnamefont {Smetanka}}, \bibinfo {author} {\bibfnamefont {T.}~\bibnamefont {Davis}}, \bibinfo {author} {\bibfnamefont {K.}~\bibnamefont {Lee}}, \bibinfo {author} {\bibfnamefont {M.~D.}\ \bibnamefont {Choi}}, \bibinfo {author} {\bibfnamefont {B.}~\bibnamefont {Boyd}}, \bibinfo {author} {\bibfnamefont {J.}~\bibnamefont {Suh}}, \bibinfo {author} {\bibfnamefont {W.}~\bibnamefont {Ha}}, \bibinfo {author} {\bibfnamefont {C.}~\bibnamefont {Jennings}}, \bibinfo {author} {\bibfnamefont {A.~S.}\ \bibnamefont {Pan}}, \bibinfo {author} {\bibfnamefont {D.~S.}\ \bibnamefont {Sanchez}}, \bibinfo {author} {\bibfnamefont {M.~D.}\ \bibnamefont
  {Reed}},\ and\ \bibinfo {author} {\bibfnamefont {J.~R.}\ \bibnamefont {Petta}},\ }\bibfield  {title} {\bibinfo {title} {Coherent control of a triangular exchange-only spin qubit},\ }\href {https://doi.org/10.1103/PhysRevApplied.22.044057} {\bibfield  {journal} {\bibinfo  {journal} {Phys. Rev. Appl.}\ }\textbf {\bibinfo {volume} {22}},\ \bibinfo {pages} {044057} (\bibinfo {year} {2024})}\BibitemShut {NoStop}%
\bibitem [{\citenamefont {Borsoi}\ \emph {et~al.}(2024)\citenamefont {Borsoi}, \citenamefont {Hendrickx}, \citenamefont {John}, \citenamefont {Meyer}, \citenamefont {Motz}, \citenamefont {van Riggelen}, \citenamefont {Sammak}, \citenamefont {de~Snoo}, \citenamefont {Scappucci},\ and\ \citenamefont {Veldhorst}}]{Borsoi2024}%
  \BibitemOpen
  \bibfield  {author} {\bibinfo {author} {\bibfnamefont {F.}~\bibnamefont {Borsoi}}, \bibinfo {author} {\bibfnamefont {N.~W.}\ \bibnamefont {Hendrickx}}, \bibinfo {author} {\bibfnamefont {V.}~\bibnamefont {John}}, \bibinfo {author} {\bibfnamefont {M.}~\bibnamefont {Meyer}}, \bibinfo {author} {\bibfnamefont {S.}~\bibnamefont {Motz}}, \bibinfo {author} {\bibfnamefont {F.}~\bibnamefont {van Riggelen}}, \bibinfo {author} {\bibfnamefont {A.}~\bibnamefont {Sammak}}, \bibinfo {author} {\bibfnamefont {S.~L.}\ \bibnamefont {de~Snoo}}, \bibinfo {author} {\bibfnamefont {G.}~\bibnamefont {Scappucci}},\ and\ \bibinfo {author} {\bibfnamefont {M.}~\bibnamefont {Veldhorst}},\ }\bibfield  {title} {\bibinfo {title} {Shared control of a 16 semiconductor quantum dot crossbar array},\ }\href {https://doi.org/10.1038/s41565-023-01491-3} {\bibfield  {journal} {\bibinfo  {journal} {Nature Nanotechnology}\ }\textbf {\bibinfo {volume} {19}},\ \bibinfo {pages} {21} (\bibinfo {year} {2024})}\BibitemShut {NoStop}%
\bibitem [{\citenamefont {Zhang}\ \emph {et~al.}(2024)\citenamefont {Zhang}, \citenamefont {Morozova}, \citenamefont {Rimbach-Russ}, \citenamefont {Jirovec}, \citenamefont {Hsiao}, \citenamefont {Fariña}, \citenamefont {Wang}, \citenamefont {Oosterhout}, \citenamefont {Sammak}, \citenamefont {Scappucci}, \citenamefont {Veldhorst},\ and\ \citenamefont {Vandersypen}}]{Zhang2024}%
  \BibitemOpen
  \bibfield  {author} {\bibinfo {author} {\bibfnamefont {X.}~\bibnamefont {Zhang}}, \bibinfo {author} {\bibfnamefont {E.}~\bibnamefont {Morozova}}, \bibinfo {author} {\bibfnamefont {M.}~\bibnamefont {Rimbach-Russ}}, \bibinfo {author} {\bibfnamefont {D.}~\bibnamefont {Jirovec}}, \bibinfo {author} {\bibfnamefont {T.-K.}\ \bibnamefont {Hsiao}}, \bibinfo {author} {\bibfnamefont {P.~C.}\ \bibnamefont {Fariña}}, \bibinfo {author} {\bibfnamefont {C.-A.}\ \bibnamefont {Wang}}, \bibinfo {author} {\bibfnamefont {S.~D.}\ \bibnamefont {Oosterhout}}, \bibinfo {author} {\bibfnamefont {A.}~\bibnamefont {Sammak}}, \bibinfo {author} {\bibfnamefont {G.}~\bibnamefont {Scappucci}}, \bibinfo {author} {\bibfnamefont {M.}~\bibnamefont {Veldhorst}},\ and\ \bibinfo {author} {\bibfnamefont {L.~M.~K.}\ \bibnamefont {Vandersypen}},\ }\bibfield  {title} {\bibinfo {title} {Universal control of four singlet–triplet qubits},\ }\bibfield  {journal} {\bibinfo  {journal} {Nature Nanotechnology}\ }\href
  {https://doi.org/10.1038/s41565-024-01817-9} {10.1038/s41565-024-01817-9} (\bibinfo {year} {2024})\BibitemShut {NoStop}%
\bibitem [{\citenamefont {Wang}\ \emph {et~al.}(2024{\natexlab{a}})\citenamefont {Wang}, \citenamefont {John}, \citenamefont {Tidjani}, \citenamefont {Yu}, \citenamefont {Ivlev}, \citenamefont {Déprez}, \citenamefont {van Riggelen-Doelman}, \citenamefont {Woods}, \citenamefont {Hendrickx}, \citenamefont {Lawrie}, \citenamefont {Stehouwer}, \citenamefont {Oosterhout}, \citenamefont {Sammak}, \citenamefont {Friesen}, \citenamefont {Scappucci}, \citenamefont {de~Snoo}, \citenamefont {Rimbach-Russ}, \citenamefont {Borsoi},\ and\ \citenamefont {Veldhorst}}]{Wang2024}%
  \BibitemOpen
  \bibfield  {author} {\bibinfo {author} {\bibfnamefont {C.-A.}\ \bibnamefont {Wang}}, \bibinfo {author} {\bibfnamefont {V.}~\bibnamefont {John}}, \bibinfo {author} {\bibfnamefont {H.}~\bibnamefont {Tidjani}}, \bibinfo {author} {\bibfnamefont {C.~X.}\ \bibnamefont {Yu}}, \bibinfo {author} {\bibfnamefont {A.~S.}\ \bibnamefont {Ivlev}}, \bibinfo {author} {\bibfnamefont {C.}~\bibnamefont {Déprez}}, \bibinfo {author} {\bibfnamefont {F.}~\bibnamefont {van Riggelen-Doelman}}, \bibinfo {author} {\bibfnamefont {B.~D.}\ \bibnamefont {Woods}}, \bibinfo {author} {\bibfnamefont {N.~W.}\ \bibnamefont {Hendrickx}}, \bibinfo {author} {\bibfnamefont {W.~I.~L.}\ \bibnamefont {Lawrie}}, \bibinfo {author} {\bibfnamefont {L.~E.~A.}\ \bibnamefont {Stehouwer}}, \bibinfo {author} {\bibfnamefont {S.~D.}\ \bibnamefont {Oosterhout}}, \bibinfo {author} {\bibfnamefont {A.}~\bibnamefont {Sammak}}, \bibinfo {author} {\bibfnamefont {M.}~\bibnamefont {Friesen}}, \bibinfo {author} {\bibfnamefont {G.}~\bibnamefont {Scappucci}}, \bibinfo
  {author} {\bibfnamefont {S.~L.}\ \bibnamefont {de~Snoo}}, \bibinfo {author} {\bibfnamefont {M.}~\bibnamefont {Rimbach-Russ}}, \bibinfo {author} {\bibfnamefont {F.}~\bibnamefont {Borsoi}},\ and\ \bibinfo {author} {\bibfnamefont {M.}~\bibnamefont {Veldhorst}},\ }\bibfield  {title} {\bibinfo {title} {Operating semiconductor quantum processors with hopping spins},\ }\href {https://doi.org/10.1126/science.ado5915} {\bibfield  {journal} {\bibinfo  {journal} {Science}\ }\textbf {\bibinfo {volume} {385}},\ \bibinfo {pages} {447} (\bibinfo {year} {2024}{\natexlab{a}})},\ \Eprint {https://arxiv.org/abs/https://www.science.org/doi/pdf/10.1126/science.ado5915} {https://www.science.org/doi/pdf/10.1126/science.ado5915} \BibitemShut {NoStop}%
\bibitem [{\citenamefont {Wang}\ \emph {et~al.}(2024{\natexlab{b}})\citenamefont {Wang}, \citenamefont {Kang}, \citenamefont {Lu}, \citenamefont {Wang}, \citenamefont {Wang}, \citenamefont {Li}, \citenamefont {Cao}, \citenamefont {Wang},\ and\ \citenamefont {Guo}}]{Wang2024b}%
  \BibitemOpen
  \bibfield  {author} {\bibinfo {author} {\bibfnamefont {N.}~\bibnamefont {Wang}}, \bibinfo {author} {\bibfnamefont {J.-M.}\ \bibnamefont {Kang}}, \bibinfo {author} {\bibfnamefont {W.-L.}\ \bibnamefont {Lu}}, \bibinfo {author} {\bibfnamefont {S.-M.}\ \bibnamefont {Wang}}, \bibinfo {author} {\bibfnamefont {Y.-J.}\ \bibnamefont {Wang}}, \bibinfo {author} {\bibfnamefont {H.-O.}\ \bibnamefont {Li}}, \bibinfo {author} {\bibfnamefont {G.}~\bibnamefont {Cao}}, \bibinfo {author} {\bibfnamefont {B.-C.}\ \bibnamefont {Wang}},\ and\ \bibinfo {author} {\bibfnamefont {G.-P.}\ \bibnamefont {Guo}},\ }\bibfield  {title} {\bibinfo {title} {Highly tunable {2D} silicon quantum dot array with coupling beyond nearest neighbors},\ }\href {https://doi.org/10.1021/acs.nanolett.4c02345} {\bibfield  {journal} {\bibinfo  {journal} {Nano Letters}\ }\textbf {\bibinfo {volume} {24}},\ \bibinfo {pages} {13126} (\bibinfo {year} {2024}{\natexlab{b}})},\ \bibinfo {note} {pMID: 39401161},\ \Eprint
  {https://arxiv.org/abs/https://doi.org/10.1021/acs.nanolett.4c02345} {https://doi.org/10.1021/acs.nanolett.4c02345} \BibitemShut {NoStop}%
\bibitem [{\citenamefont {Ha}\ \emph {et~al.}(2025)\citenamefont {Ha}, \citenamefont {Acuna}, \citenamefont {Raach}, \citenamefont {Bloom}, \citenamefont {Brecht}, \citenamefont {Chappell}, \citenamefont {Choi}, \citenamefont {Christensen}, \citenamefont {Counts}, \citenamefont {Daprano}, \citenamefont {Dodson}, \citenamefont {Eng}, \citenamefont {Fialkow}, \citenamefont {Garcia}, \citenamefont {Ha}, \citenamefont {Harris}, \citenamefont {Holman}, \citenamefont {Khalaf}, \citenamefont {Matten}, \citenamefont {Peterson}, \citenamefont {Plesha}, \citenamefont {Ruiz}, \citenamefont {Smith}, \citenamefont {Thomas}, \citenamefont {Whiteley}, \citenamefont {Ladd}, \citenamefont {Jura}, \citenamefont {Rakher},\ and\ \citenamefont {Borselli}}]{Ha2025}%
  \BibitemOpen
  \bibfield  {author} {\bibinfo {author} {\bibfnamefont {S.~D.}\ \bibnamefont {Ha}}, \bibinfo {author} {\bibfnamefont {E.}~\bibnamefont {Acuna}}, \bibinfo {author} {\bibfnamefont {K.}~\bibnamefont {Raach}}, \bibinfo {author} {\bibfnamefont {Z.~T.}\ \bibnamefont {Bloom}}, \bibinfo {author} {\bibfnamefont {T.~L.}\ \bibnamefont {Brecht}}, \bibinfo {author} {\bibfnamefont {J.~M.}\ \bibnamefont {Chappell}}, \bibinfo {author} {\bibfnamefont {M.~D.}\ \bibnamefont {Choi}}, \bibinfo {author} {\bibfnamefont {J.~E.}\ \bibnamefont {Christensen}}, \bibinfo {author} {\bibfnamefont {I.~T.}\ \bibnamefont {Counts}}, \bibinfo {author} {\bibfnamefont {D.}~\bibnamefont {Daprano}}, \bibinfo {author} {\bibfnamefont {J.~P.}\ \bibnamefont {Dodson}}, \bibinfo {author} {\bibfnamefont {K.}~\bibnamefont {Eng}}, \bibinfo {author} {\bibfnamefont {D.~J.}\ \bibnamefont {Fialkow}}, \bibinfo {author} {\bibfnamefont {C.~A.~C.}\ \bibnamefont {Garcia}}, \bibinfo {author} {\bibfnamefont {W.}~\bibnamefont {Ha}}, \bibinfo {author} {\bibfnamefont
  {T.~R.~B.}\ \bibnamefont {Harris}}, \bibinfo {author} {\bibfnamefont {N.}~\bibnamefont {Holman}}, \bibinfo {author} {\bibfnamefont {I.}~\bibnamefont {Khalaf}}, \bibinfo {author} {\bibfnamefont {J.~W.}\ \bibnamefont {Matten}}, \bibinfo {author} {\bibfnamefont {C.~A.}\ \bibnamefont {Peterson}}, \bibinfo {author} {\bibfnamefont {C.~E.}\ \bibnamefont {Plesha}}, \bibinfo {author} {\bibfnamefont {M.~J.}\ \bibnamefont {Ruiz}}, \bibinfo {author} {\bibfnamefont {A.}~\bibnamefont {Smith}}, \bibinfo {author} {\bibfnamefont {B.~J.}\ \bibnamefont {Thomas}}, \bibinfo {author} {\bibfnamefont {S.~J.}\ \bibnamefont {Whiteley}}, \bibinfo {author} {\bibfnamefont {T.~D.}\ \bibnamefont {Ladd}}, \bibinfo {author} {\bibfnamefont {M.~P.}\ \bibnamefont {Jura}}, \bibinfo {author} {\bibfnamefont {M.~T.}\ \bibnamefont {Rakher}},\ and\ \bibinfo {author} {\bibfnamefont {M.~G.}\ \bibnamefont {Borselli}},\ }\bibfield  {title} {\bibinfo {title} {Two-dimensional {Si} spin qubit arrays with multilevel interconnects},\ }\href
  {https://arxiv.org/abs/2502.08861} {\bibfield  {journal} {\bibinfo  {journal} {arXiv:2502.08861}\ } (\bibinfo {year} {2025})}\BibitemShut {NoStop}%
\bibitem [{\citenamefont {Buterakos}\ and\ \citenamefont {Das~Sarma}(2021)}]{Buterakos2021}%
  \BibitemOpen
  \bibfield  {author} {\bibinfo {author} {\bibfnamefont {D.}~\bibnamefont {Buterakos}}\ and\ \bibinfo {author} {\bibfnamefont {S.}~\bibnamefont {Das~Sarma}},\ }\bibfield  {title} {\bibinfo {title} {Spin-valley qubit dynamics in exchange-coupled silicon quantum dots},\ }\href {https://doi.org/10.1103/PRXQuantum.2.040358} {\bibfield  {journal} {\bibinfo  {journal} {PRX Quantum}\ }\textbf {\bibinfo {volume} {2}},\ \bibinfo {pages} {040358} (\bibinfo {year} {2021})}\BibitemShut {NoStop}%
\bibitem [{\citenamefont {Zwanenburg}\ \emph {et~al.}(2013)\citenamefont {Zwanenburg}, \citenamefont {Dzurak}, \citenamefont {Morello}, \citenamefont {Simmons}, \citenamefont {Hollenberg}, \citenamefont {Klimeck}, \citenamefont {Rogge}, \citenamefont {Coppersmith},\ and\ \citenamefont {Eriksson}}]{Zwanenburg2013}%
  \BibitemOpen
  \bibfield  {author} {\bibinfo {author} {\bibfnamefont {F.~A.}\ \bibnamefont {Zwanenburg}}, \bibinfo {author} {\bibfnamefont {A.~S.}\ \bibnamefont {Dzurak}}, \bibinfo {author} {\bibfnamefont {A.}~\bibnamefont {Morello}}, \bibinfo {author} {\bibfnamefont {M.~Y.}\ \bibnamefont {Simmons}}, \bibinfo {author} {\bibfnamefont {L.~C.~L.}\ \bibnamefont {Hollenberg}}, \bibinfo {author} {\bibfnamefont {G.}~\bibnamefont {Klimeck}}, \bibinfo {author} {\bibfnamefont {S.}~\bibnamefont {Rogge}}, \bibinfo {author} {\bibfnamefont {S.~N.}\ \bibnamefont {Coppersmith}},\ and\ \bibinfo {author} {\bibfnamefont {M.~A.}\ \bibnamefont {Eriksson}},\ }\bibfield  {title} {\bibinfo {title} {Silicon quantum electronics},\ }\href {https://doi.org/10.1103/RevModPhys.85.961} {\bibfield  {journal} {\bibinfo  {journal} {Rev. Mod. Phys.}\ }\textbf {\bibinfo {volume} {85}},\ \bibinfo {pages} {961} (\bibinfo {year} {2013})}\BibitemShut {NoStop}%
\bibitem [{\citenamefont {Gyure}\ \emph {et~al.}(2021)\citenamefont {Gyure}, \citenamefont {Kiselev}, \citenamefont {Ross}, \citenamefont {Rahman},\ and\ \citenamefont {de~Walle}}]{Gyure2021}%
  \BibitemOpen
  \bibfield  {author} {\bibinfo {author} {\bibfnamefont {M.~F.}\ \bibnamefont {Gyure}}, \bibinfo {author} {\bibfnamefont {A.~A.}\ \bibnamefont {Kiselev}}, \bibinfo {author} {\bibfnamefont {R.~S.}\ \bibnamefont {Ross}}, \bibinfo {author} {\bibfnamefont {R.}~\bibnamefont {Rahman}},\ and\ \bibinfo {author} {\bibfnamefont {C.~G.~V.}\ \bibnamefont {de~Walle}},\ }\bibfield  {title} {\bibinfo {title} {Materials and device simulations for silicon qubit design and optimization},\ }\href {https://doi.org/10.1557/s43577-021-00140-1} {\bibfield  {journal} {\bibinfo  {journal} {MRS Bulletin}\ }\textbf {\bibinfo {volume} {46}},\ \bibinfo {pages} {634} (\bibinfo {year} {2021})}\BibitemShut {NoStop}%
\bibitem [{\citenamefont {Wuetz}\ \emph {et~al.}(2022)\citenamefont {Wuetz}, \citenamefont {Losert}, \citenamefont {Koelling}, \citenamefont {Stehouwer}, \citenamefont {Zwerver}, \citenamefont {Philips}, \citenamefont {Mądzik}, \citenamefont {Xue}, \citenamefont {Zheng}, \citenamefont {Lodari}, \citenamefont {Amitonov}, \citenamefont {Samkharadze}, \citenamefont {Sammak}, \citenamefont {Vandersypen}, \citenamefont {Rahman}, \citenamefont {Coppersmith}, \citenamefont {Moutanabbir}, \citenamefont {Friesen},\ and\ \citenamefont {Scappucci}}]{Wuetz2021}%
  \BibitemOpen
  \bibfield  {author} {\bibinfo {author} {\bibfnamefont {B.~P.}\ \bibnamefont {Wuetz}}, \bibinfo {author} {\bibfnamefont {M.~P.}\ \bibnamefont {Losert}}, \bibinfo {author} {\bibfnamefont {S.}~\bibnamefont {Koelling}}, \bibinfo {author} {\bibfnamefont {L.~E.~A.}\ \bibnamefont {Stehouwer}}, \bibinfo {author} {\bibfnamefont {A.-M.~J.}\ \bibnamefont {Zwerver}}, \bibinfo {author} {\bibfnamefont {S.~G.~J.}\ \bibnamefont {Philips}}, \bibinfo {author} {\bibfnamefont {M.~T.}\ \bibnamefont {Mądzik}}, \bibinfo {author} {\bibfnamefont {X.}~\bibnamefont {Xue}}, \bibinfo {author} {\bibfnamefont {G.}~\bibnamefont {Zheng}}, \bibinfo {author} {\bibfnamefont {M.}~\bibnamefont {Lodari}}, \bibinfo {author} {\bibfnamefont {S.~V.}\ \bibnamefont {Amitonov}}, \bibinfo {author} {\bibfnamefont {N.}~\bibnamefont {Samkharadze}}, \bibinfo {author} {\bibfnamefont {A.}~\bibnamefont {Sammak}}, \bibinfo {author} {\bibfnamefont {L.~M.~K.}\ \bibnamefont {Vandersypen}}, \bibinfo {author} {\bibfnamefont {R.}~\bibnamefont {Rahman}}, \bibinfo
  {author} {\bibfnamefont {S.~N.}\ \bibnamefont {Coppersmith}}, \bibinfo {author} {\bibfnamefont {O.}~\bibnamefont {Moutanabbir}}, \bibinfo {author} {\bibfnamefont {M.}~\bibnamefont {Friesen}},\ and\ \bibinfo {author} {\bibfnamefont {G.}~\bibnamefont {Scappucci}},\ }\bibfield  {title} {\bibinfo {title} {Atomic fluctuations lifting the energy degeneracy in {Si/SiGe} quantum dots},\ }\href {https://doi.org/10.1038/s41467-022-35458-0} {\bibfield  {journal} {\bibinfo  {journal} {Nature Communications}\ }\textbf {\bibinfo {volume} {13}},\ \bibinfo {pages} {7730} (\bibinfo {year} {2022})}\BibitemShut {NoStop}%
\bibitem [{\citenamefont {Losert}\ \emph {et~al.}(2023)\citenamefont {Losert}, \citenamefont {Eriksson}, \citenamefont {Joynt}, \citenamefont {Rahman}, \citenamefont {Scappucci}, \citenamefont {Coppersmith},\ and\ \citenamefont {Friesen}}]{Losert2023}%
  \BibitemOpen
  \bibfield  {author} {\bibinfo {author} {\bibfnamefont {M.~P.}\ \bibnamefont {Losert}}, \bibinfo {author} {\bibfnamefont {M.~A.}\ \bibnamefont {Eriksson}}, \bibinfo {author} {\bibfnamefont {R.}~\bibnamefont {Joynt}}, \bibinfo {author} {\bibfnamefont {R.}~\bibnamefont {Rahman}}, \bibinfo {author} {\bibfnamefont {G.}~\bibnamefont {Scappucci}}, \bibinfo {author} {\bibfnamefont {S.~N.}\ \bibnamefont {Coppersmith}},\ and\ \bibinfo {author} {\bibfnamefont {M.}~\bibnamefont {Friesen}},\ }\bibfield  {title} {\bibinfo {title} {Practical strategies for enhancing the valley splitting in {Si/SiGe} quantum wells},\ }\href {https://doi.org/10.1103/PhysRevB.108.125405} {\bibfield  {journal} {\bibinfo  {journal} {Phys. Rev. B}\ }\textbf {\bibinfo {volume} {108}},\ \bibinfo {pages} {125405} (\bibinfo {year} {2023})}\BibitemShut {NoStop}%
\bibitem [{\citenamefont {McJunkin}\ \emph {et~al.}(2022)\citenamefont {McJunkin}, \citenamefont {Harpt}, \citenamefont {Feng}, \citenamefont {Losert}, \citenamefont {Rahman}, \citenamefont {Dodson}, \citenamefont {Wolfe}, \citenamefont {Savage}, \citenamefont {Lagally}, \citenamefont {Coppersmith}, \citenamefont {Friesen}, \citenamefont {Joynt},\ and\ \citenamefont {Eriksson}}]{McJunkin2021}%
  \BibitemOpen
  \bibfield  {author} {\bibinfo {author} {\bibfnamefont {T.}~\bibnamefont {McJunkin}}, \bibinfo {author} {\bibfnamefont {B.}~\bibnamefont {Harpt}}, \bibinfo {author} {\bibfnamefont {Y.}~\bibnamefont {Feng}}, \bibinfo {author} {\bibfnamefont {M.~P.}\ \bibnamefont {Losert}}, \bibinfo {author} {\bibfnamefont {R.}~\bibnamefont {Rahman}}, \bibinfo {author} {\bibfnamefont {J.~P.}\ \bibnamefont {Dodson}}, \bibinfo {author} {\bibfnamefont {M.~A.}\ \bibnamefont {Wolfe}}, \bibinfo {author} {\bibfnamefont {D.~E.}\ \bibnamefont {Savage}}, \bibinfo {author} {\bibfnamefont {M.~G.}\ \bibnamefont {Lagally}}, \bibinfo {author} {\bibfnamefont {S.~N.}\ \bibnamefont {Coppersmith}}, \bibinfo {author} {\bibfnamefont {M.}~\bibnamefont {Friesen}}, \bibinfo {author} {\bibfnamefont {R.}~\bibnamefont {Joynt}},\ and\ \bibinfo {author} {\bibfnamefont {M.~A.}\ \bibnamefont {Eriksson}},\ }\bibfield  {title} {\bibinfo {title} {{SiGe} quantum wells with oscillating {Ge} concentrations for quantum dot qubits},\ }\href
  {https://doi.org/10.1038/s41467-022-35510-z} {\bibfield  {journal} {\bibinfo  {journal} {Nature Communications}\ }\textbf {\bibinfo {volume} {13}},\ \bibinfo {pages} {7777} (\bibinfo {year} {2022})}\BibitemShut {NoStop}%
\bibitem [{\citenamefont {Lima}\ and\ \citenamefont {Burkard}(2023)}]{Lima2023a}%
  \BibitemOpen
  \bibfield  {author} {\bibinfo {author} {\bibfnamefont {J.~R.~F.}\ \bibnamefont {Lima}}\ and\ \bibinfo {author} {\bibfnamefont {G.}~\bibnamefont {Burkard}},\ }\bibfield  {title} {\bibinfo {title} {Interface and electromagnetic effects in the valley splitting of {Si} quantum dots},\ }\href {https://doi.org/10.1088/2633-4356/acd743} {\bibfield  {journal} {\bibinfo  {journal} {Materials for Quantum Technology}\ }\textbf {\bibinfo {volume} {3}},\ \bibinfo {pages} {025004} (\bibinfo {year} {2023})}\BibitemShut {NoStop}%
\bibitem [{\citenamefont {Lima}\ and\ \citenamefont {Burkard}(2024)}]{Lima2023b}%
  \BibitemOpen
  \bibfield  {author} {\bibinfo {author} {\bibfnamefont {J.~R.~F.}\ \bibnamefont {Lima}}\ and\ \bibinfo {author} {\bibfnamefont {G.}~\bibnamefont {Burkard}},\ }\bibfield  {title} {\bibinfo {title} {Valley splitting depending on the size and location of a silicon quantum dot},\ }\href {https://doi.org/10.1103/PhysRevMaterials.8.036202} {\bibfield  {journal} {\bibinfo  {journal} {Phys. Rev. Mater.}\ }\textbf {\bibinfo {volume} {8}},\ \bibinfo {pages} {036202} (\bibinfo {year} {2024})}\BibitemShut {NoStop}%
\bibitem [{\citenamefont {Nagaoka}(1966)}]{Nagaoka1966}%
  \BibitemOpen
  \bibfield  {author} {\bibinfo {author} {\bibfnamefont {Y.}~\bibnamefont {Nagaoka}},\ }\bibfield  {title} {\bibinfo {title} {Ferromagnetism in a narrow, almost half-filled $s$ band},\ }\href {https://doi.org/10.1103/PhysRev.147.392} {\bibfield  {journal} {\bibinfo  {journal} {Phys. Rev.}\ }\textbf {\bibinfo {volume} {147}},\ \bibinfo {pages} {392} (\bibinfo {year} {1966})}\BibitemShut {NoStop}%
\bibitem [{\citenamefont {Tasaki}(1998)}]{Tasaki1998}%
  \BibitemOpen
  \bibfield  {author} {\bibinfo {author} {\bibfnamefont {H.}~\bibnamefont {Tasaki}},\ }\bibfield  {title} {\bibinfo {title} {The {Hubbard} model - an introduction and selected rigorous results},\ }\href {https://doi.org/10.1088/0953-8984/10/20/004} {\bibfield  {journal} {\bibinfo  {journal} {Journal of Physics: Condensed Matter}\ }\textbf {\bibinfo {volume} {10}},\ \bibinfo {pages} {4353} (\bibinfo {year} {1998})}\BibitemShut {NoStop}%
\bibitem [{\citenamefont {Dehollain}\ \emph {et~al.}(2020)\citenamefont {Dehollain}, \citenamefont {Mukhopadhyay}, \citenamefont {Michal}, \citenamefont {Wang}, \citenamefont {Wunsch}, \citenamefont {Reichl}, \citenamefont {Wegscheider}, \citenamefont {Rudner}, \citenamefont {Demler},\ and\ \citenamefont {Vandersypen}}]{Dehollain2020}%
  \BibitemOpen
  \bibfield  {author} {\bibinfo {author} {\bibfnamefont {J.~P.}\ \bibnamefont {Dehollain}}, \bibinfo {author} {\bibfnamefont {U.}~\bibnamefont {Mukhopadhyay}}, \bibinfo {author} {\bibfnamefont {V.~P.}\ \bibnamefont {Michal}}, \bibinfo {author} {\bibfnamefont {Y.}~\bibnamefont {Wang}}, \bibinfo {author} {\bibfnamefont {B.}~\bibnamefont {Wunsch}}, \bibinfo {author} {\bibfnamefont {C.}~\bibnamefont {Reichl}}, \bibinfo {author} {\bibfnamefont {W.}~\bibnamefont {Wegscheider}}, \bibinfo {author} {\bibfnamefont {M.~S.}\ \bibnamefont {Rudner}}, \bibinfo {author} {\bibfnamefont {E.}~\bibnamefont {Demler}},\ and\ \bibinfo {author} {\bibfnamefont {L.~M.~K.}\ \bibnamefont {Vandersypen}},\ }\bibfield  {title} {\bibinfo {title} {Nagaoka ferromagnetism observed in a quantum dot plaquette},\ }\href {https://doi.org/10.1038/s41586-020-2051-0} {\bibfield  {journal} {\bibinfo  {journal} {Nature}\ }\textbf {\bibinfo {volume} {579}},\ \bibinfo {pages} {528} (\bibinfo {year} {2020})}\BibitemShut {NoStop}%
\bibitem [{\citenamefont {Haldane}(1983)}]{Haldane1983}%
  \BibitemOpen
  \bibfield  {author} {\bibinfo {author} {\bibfnamefont {F.}~\bibnamefont {Haldane}},\ }\bibfield  {title} {\bibinfo {title} {Continuum dynamics of the {1-D Heisenberg} antiferromagnet: Identification with the {O}(3) nonlinear sigma model},\ }\href {https://doi.org/https://doi.org/10.1016/0375-9601(83)90631-X} {\bibfield  {journal} {\bibinfo  {journal} {Physics Letters A}\ }\textbf {\bibinfo {volume} {93}},\ \bibinfo {pages} {464} (\bibinfo {year} {1983})}\BibitemShut {NoStop}%
\bibitem [{\citenamefont {Shim}\ \emph {et~al.}(2010)\citenamefont {Shim}, \citenamefont {Sharma}, \citenamefont {Hsieh},\ and\ \citenamefont {Hawrylak}}]{Shim2010}%
  \BibitemOpen
  \bibfield  {author} {\bibinfo {author} {\bibfnamefont {Y.-P.}\ \bibnamefont {Shim}}, \bibinfo {author} {\bibfnamefont {A.}~\bibnamefont {Sharma}}, \bibinfo {author} {\bibfnamefont {C.-Y.}\ \bibnamefont {Hsieh}},\ and\ \bibinfo {author} {\bibfnamefont {P.}~\bibnamefont {Hawrylak}},\ }\bibfield  {title} {\bibinfo {title} {Artificial {Haldane} gap material on a semiconductor chip},\ }\href {https://doi.org/https://doi.org/10.1016/j.ssc.2010.08.002} {\bibfield  {journal} {\bibinfo  {journal} {Solid State Communications}\ }\textbf {\bibinfo {volume} {150}},\ \bibinfo {pages} {2065} (\bibinfo {year} {2010})}\BibitemShut {NoStop}%
\bibitem [{\citenamefont {Sugimoto}\ \emph {et~al.}(2020)\citenamefont {Sugimoto}, \citenamefont {Morita},\ and\ \citenamefont {Tohyama}}]{Sugimoto2020}%
  \BibitemOpen
  \bibfield  {author} {\bibinfo {author} {\bibfnamefont {T.}~\bibnamefont {Sugimoto}}, \bibinfo {author} {\bibfnamefont {K.}~\bibnamefont {Morita}},\ and\ \bibinfo {author} {\bibfnamefont {T.}~\bibnamefont {Tohyama}},\ }\bibfield  {title} {\bibinfo {title} {Cluster-based {Haldane} states in spin-1/2 cluster chains},\ }\href {https://doi.org/10.1103/PhysRevResearch.2.023420} {\bibfield  {journal} {\bibinfo  {journal} {Phys. Rev. Res.}\ }\textbf {\bibinfo {volume} {2}},\ \bibinfo {pages} {023420} (\bibinfo {year} {2020})}\BibitemShut {NoStop}%
\bibitem [{\citenamefont {Catarina}\ and\ \citenamefont {Fern\'andez-Rossier}(2022)}]{Catarina2022}%
  \BibitemOpen
  \bibfield  {author} {\bibinfo {author} {\bibfnamefont {G.}~\bibnamefont {Catarina}}\ and\ \bibinfo {author} {\bibfnamefont {J.}~\bibnamefont {Fern\'andez-Rossier}},\ }\bibfield  {title} {\bibinfo {title} {Hubbard model for spin-1 {Haldane} chains},\ }\href {https://doi.org/10.1103/PhysRevB.105.L081116} {\bibfield  {journal} {\bibinfo  {journal} {Phys. Rev. B}\ }\textbf {\bibinfo {volume} {105}},\ \bibinfo {pages} {L081116} (\bibinfo {year} {2022})}\BibitemShut {NoStop}%
\bibitem [{\citenamefont {Baran}\ and\ \citenamefont {Paaske}(2024)}]{Baran2024}%
  \BibitemOpen
  \bibfield  {author} {\bibinfo {author} {\bibfnamefont {V.~V.}\ \bibnamefont {Baran}}\ and\ \bibinfo {author} {\bibfnamefont {J.}~\bibnamefont {Paaske}},\ }\bibfield  {title} {\bibinfo {title} {Spin-1 {Haldane} chains of superconductor-semiconductor hybrids},\ }\href {https://doi.org/10.1103/PhysRevB.110.064503} {\bibfield  {journal} {\bibinfo  {journal} {Phys. Rev. B}\ }\textbf {\bibinfo {volume} {110}},\ \bibinfo {pages} {064503} (\bibinfo {year} {2024})}\BibitemShut {NoStop}%
\bibitem [{\citenamefont {Manalo}\ \emph {et~al.}(2024)\citenamefont {Manalo}, \citenamefont {Miravet},\ and\ \citenamefont {Hawrylak}}]{Manalo2024}%
  \BibitemOpen
  \bibfield  {author} {\bibinfo {author} {\bibfnamefont {J.}~\bibnamefont {Manalo}}, \bibinfo {author} {\bibfnamefont {D.}~\bibnamefont {Miravet}},\ and\ \bibinfo {author} {\bibfnamefont {P.}~\bibnamefont {Hawrylak}},\ }\bibfield  {title} {\bibinfo {title} {Microscopic design of a synthetic spin-1 chain in an {InAsP} quantum dot array},\ }\href {https://doi.org/10.1103/PhysRevB.109.085112} {\bibfield  {journal} {\bibinfo  {journal} {Phys. Rev. B}\ }\textbf {\bibinfo {volume} {109}},\ \bibinfo {pages} {085112} (\bibinfo {year} {2024})}\BibitemShut {NoStop}%
\bibitem [{\citenamefont {Cooper}\ \emph {et~al.}(2019)\citenamefont {Cooper}, \citenamefont {Dalibard},\ and\ \citenamefont {Spielman}}]{Cooper2019}%
  \BibitemOpen
  \bibfield  {author} {\bibinfo {author} {\bibfnamefont {N.~R.}\ \bibnamefont {Cooper}}, \bibinfo {author} {\bibfnamefont {J.}~\bibnamefont {Dalibard}},\ and\ \bibinfo {author} {\bibfnamefont {I.~B.}\ \bibnamefont {Spielman}},\ }\bibfield  {title} {\bibinfo {title} {Topological bands for ultracold atoms},\ }\href {https://doi.org/10.1103/RevModPhys.91.015005} {\bibfield  {journal} {\bibinfo  {journal} {Rev. Mod. Phys.}\ }\textbf {\bibinfo {volume} {91}},\ \bibinfo {pages} {015005} (\bibinfo {year} {2019})}\BibitemShut {NoStop}%
\bibitem [{\citenamefont {Dalibard}\ \emph {et~al.}(2011)\citenamefont {Dalibard}, \citenamefont {Gerbier}, \citenamefont {Juzeli\ifmmode~\bar{u}\else \={u}\fi{}nas},\ and\ \citenamefont {\"Ohberg}}]{Dalibard2011}%
  \BibitemOpen
  \bibfield  {author} {\bibinfo {author} {\bibfnamefont {J.}~\bibnamefont {Dalibard}}, \bibinfo {author} {\bibfnamefont {F.}~\bibnamefont {Gerbier}}, \bibinfo {author} {\bibfnamefont {G.}~\bibnamefont {Juzeli\ifmmode~\bar{u}\else \={u}\fi{}nas}},\ and\ \bibinfo {author} {\bibfnamefont {P.}~\bibnamefont {\"Ohberg}},\ }\bibfield  {title} {\bibinfo {title} {Colloquium: Artificial gauge potentials for neutral atoms},\ }\href {https://doi.org/10.1103/RevModPhys.83.1523} {\bibfield  {journal} {\bibinfo  {journal} {Rev. Mod. Phys.}\ }\textbf {\bibinfo {volume} {83}},\ \bibinfo {pages} {1523} (\bibinfo {year} {2011})}\BibitemShut {NoStop}%
\bibitem [{\citenamefont {Ozawa}\ \emph {et~al.}(2019)\citenamefont {Ozawa}, \citenamefont {Price}, \citenamefont {Amo}, \citenamefont {Goldman}, \citenamefont {Hafezi}, \citenamefont {Lu}, \citenamefont {Rechtsman}, \citenamefont {Schuster}, \citenamefont {Simon}, \citenamefont {Zilberberg},\ and\ \citenamefont {Carusotto}}]{Ozawa2019}%
  \BibitemOpen
  \bibfield  {author} {\bibinfo {author} {\bibfnamefont {T.}~\bibnamefont {Ozawa}}, \bibinfo {author} {\bibfnamefont {H.~M.}\ \bibnamefont {Price}}, \bibinfo {author} {\bibfnamefont {A.}~\bibnamefont {Amo}}, \bibinfo {author} {\bibfnamefont {N.}~\bibnamefont {Goldman}}, \bibinfo {author} {\bibfnamefont {M.}~\bibnamefont {Hafezi}}, \bibinfo {author} {\bibfnamefont {L.}~\bibnamefont {Lu}}, \bibinfo {author} {\bibfnamefont {M.~C.}\ \bibnamefont {Rechtsman}}, \bibinfo {author} {\bibfnamefont {D.}~\bibnamefont {Schuster}}, \bibinfo {author} {\bibfnamefont {J.}~\bibnamefont {Simon}}, \bibinfo {author} {\bibfnamefont {O.}~\bibnamefont {Zilberberg}},\ and\ \bibinfo {author} {\bibfnamefont {I.}~\bibnamefont {Carusotto}},\ }\bibfield  {title} {\bibinfo {title} {Topological photonics},\ }\href {https://doi.org/10.1103/RevModPhys.91.015006} {\bibfield  {journal} {\bibinfo  {journal} {Rev. Mod. Phys.}\ }\textbf {\bibinfo {volume} {91}},\ \bibinfo {pages} {015006} (\bibinfo {year} {2019})}\BibitemShut {NoStop}%
\bibitem [{\citenamefont {Yang}\ \emph {et~al.}(2015)\citenamefont {Yang}, \citenamefont {Gao}, \citenamefont {Shi}, \citenamefont {Lin}, \citenamefont {Gao}, \citenamefont {Chong},\ and\ \citenamefont {Zhang}}]{Yang2015}%
  \BibitemOpen
  \bibfield  {author} {\bibinfo {author} {\bibfnamefont {Z.}~\bibnamefont {Yang}}, \bibinfo {author} {\bibfnamefont {F.}~\bibnamefont {Gao}}, \bibinfo {author} {\bibfnamefont {X.}~\bibnamefont {Shi}}, \bibinfo {author} {\bibfnamefont {X.}~\bibnamefont {Lin}}, \bibinfo {author} {\bibfnamefont {Z.}~\bibnamefont {Gao}}, \bibinfo {author} {\bibfnamefont {Y.}~\bibnamefont {Chong}},\ and\ \bibinfo {author} {\bibfnamefont {B.}~\bibnamefont {Zhang}},\ }\bibfield  {title} {\bibinfo {title} {Topological acoustics},\ }\href {https://doi.org/10.1103/PhysRevLett.114.114301} {\bibfield  {journal} {\bibinfo  {journal} {Phys. Rev. Lett.}\ }\textbf {\bibinfo {volume} {114}},\ \bibinfo {pages} {114301} (\bibinfo {year} {2015})}\BibitemShut {NoStop}%
\bibitem [{\citenamefont {Xue}\ \emph {et~al.}(2022{\natexlab{b}})\citenamefont {Xue}, \citenamefont {Yang},\ and\ \citenamefont {Zhang}}]{Xue2022b}%
  \BibitemOpen
  \bibfield  {author} {\bibinfo {author} {\bibfnamefont {H.}~\bibnamefont {Xue}}, \bibinfo {author} {\bibfnamefont {Y.}~\bibnamefont {Yang}},\ and\ \bibinfo {author} {\bibfnamefont {B.}~\bibnamefont {Zhang}},\ }\bibfield  {title} {\bibinfo {title} {Topological acoustics},\ }\href {https://doi.org/10.1038/s41578-022-00465-6} {\bibfield  {journal} {\bibinfo  {journal} {Nature Reviews Materials}\ }\textbf {\bibinfo {volume} {7}},\ \bibinfo {pages} {974} (\bibinfo {year} {2022}{\natexlab{b}})}\BibitemShut {NoStop}%
\bibitem [{\citenamefont {Wang}\ \emph {et~al.}(2024{\natexlab{c}})\citenamefont {Wang}, \citenamefont {Zhang}, \citenamefont {Tang},\ and\ \citenamefont {Wang}}]{Wang2024c}%
  \BibitemOpen
  \bibfield  {author} {\bibinfo {author} {\bibfnamefont {W.}~\bibnamefont {Wang}}, \bibinfo {author} {\bibfnamefont {Z.}~\bibnamefont {Zhang}}, \bibinfo {author} {\bibfnamefont {G.-X.}\ \bibnamefont {Tang}},\ and\ \bibinfo {author} {\bibfnamefont {T.}~\bibnamefont {Wang}},\ }\bibfield  {title} {\bibinfo {title} {Floquet engineering tunable periodic gauge fields and simulating real topological phases in a cold-alkaline-earth-metal-atom optical lattice},\ }\href {https://doi.org/10.1103/PhysRevA.110.023308} {\bibfield  {journal} {\bibinfo  {journal} {Phys. Rev. A}\ }\textbf {\bibinfo {volume} {110}},\ \bibinfo {pages} {023308} (\bibinfo {year} {2024}{\natexlab{c}})}\BibitemShut {NoStop}%
\bibitem [{\citenamefont {Chen}\ \emph {et~al.}(2023)\citenamefont {Chen}, \citenamefont {Zhang}, \citenamefont {Yang},\ and\ \citenamefont {Zhao}}]{Chen2023}%
  \BibitemOpen
  \bibfield  {author} {\bibinfo {author} {\bibfnamefont {Z.~Y.}\ \bibnamefont {Chen}}, \bibinfo {author} {\bibfnamefont {Z.}~\bibnamefont {Zhang}}, \bibinfo {author} {\bibfnamefont {S.~A.}\ \bibnamefont {Yang}},\ and\ \bibinfo {author} {\bibfnamefont {Y.~X.}\ \bibnamefont {Zhao}},\ }\bibfield  {title} {\bibinfo {title} {Classification of time-reversal-invariant crystals with gauge structures},\ }\href {https://doi.org/10.1038/s41467-023-36447-7} {\bibfield  {journal} {\bibinfo  {journal} {Nature Communications}\ }\textbf {\bibinfo {volume} {14}},\ \bibinfo {pages} {743} (\bibinfo {year} {2023})}\BibitemShut {NoStop}%
\bibitem [{\citenamefont {Xue}\ \emph {et~al.}(2022{\natexlab{c}})\citenamefont {Xue}, \citenamefont {Wang}, \citenamefont {Huang}, \citenamefont {Cheng}, \citenamefont {Yu}, \citenamefont {Foo}, \citenamefont {Zhao}, \citenamefont {Yang},\ and\ \citenamefont {Zhang}}]{Xue2022c}%
  \BibitemOpen
  \bibfield  {author} {\bibinfo {author} {\bibfnamefont {H.}~\bibnamefont {Xue}}, \bibinfo {author} {\bibfnamefont {Z.}~\bibnamefont {Wang}}, \bibinfo {author} {\bibfnamefont {Y.-X.}\ \bibnamefont {Huang}}, \bibinfo {author} {\bibfnamefont {Z.}~\bibnamefont {Cheng}}, \bibinfo {author} {\bibfnamefont {L.}~\bibnamefont {Yu}}, \bibinfo {author} {\bibfnamefont {Y.~X.}\ \bibnamefont {Foo}}, \bibinfo {author} {\bibfnamefont {Y.~X.}\ \bibnamefont {Zhao}}, \bibinfo {author} {\bibfnamefont {S.~A.}\ \bibnamefont {Yang}},\ and\ \bibinfo {author} {\bibfnamefont {B.}~\bibnamefont {Zhang}},\ }\bibfield  {title} {\bibinfo {title} {Projectively enriched symmetry and topology in acoustic crystals},\ }\href {https://doi.org/10.1103/PhysRevLett.128.116802} {\bibfield  {journal} {\bibinfo  {journal} {Phys. Rev. Lett.}\ }\textbf {\bibinfo {volume} {128}},\ \bibinfo {pages} {116802} (\bibinfo {year} {2022}{\natexlab{c}})}\BibitemShut {NoStop}%
\bibitem [{\citenamefont {Shao}\ \emph {et~al.}(2023)\citenamefont {Shao}, \citenamefont {Chen}, \citenamefont {Wang}, \citenamefont {Yang},\ and\ \citenamefont {Zhao}}]{Shao2023}%
  \BibitemOpen
  \bibfield  {author} {\bibinfo {author} {\bibfnamefont {L.}~\bibnamefont {Shao}}, \bibinfo {author} {\bibfnamefont {Z.}~\bibnamefont {Chen}}, \bibinfo {author} {\bibfnamefont {K.}~\bibnamefont {Wang}}, \bibinfo {author} {\bibfnamefont {S.~A.}\ \bibnamefont {Yang}},\ and\ \bibinfo {author} {\bibfnamefont {Y.}~\bibnamefont {Zhao}},\ }\bibfield  {title} {\bibinfo {title} {Spinless mirror {Chern} insulator from projective symmetry algebra},\ }\href {https://doi.org/10.1103/PhysRevB.108.205126} {\bibfield  {journal} {\bibinfo  {journal} {Phys. Rev. B}\ }\textbf {\bibinfo {volume} {108}},\ \bibinfo {pages} {205126} (\bibinfo {year} {2023})}\BibitemShut {NoStop}%
\bibitem [{\citenamefont {Jiang}\ \emph {et~al.}(2024)\citenamefont {Jiang}, \citenamefont {Chen}, \citenamefont {Yue}, \citenamefont {Rui}, \citenamefont {Zhu}, \citenamefont {Yang},\ and\ \citenamefont {Zhao}}]{Jiang2024}%
  \BibitemOpen
  \bibfield  {author} {\bibinfo {author} {\bibfnamefont {G.}~\bibnamefont {Jiang}}, \bibinfo {author} {\bibfnamefont {Z.~Y.}\ \bibnamefont {Chen}}, \bibinfo {author} {\bibfnamefont {S.~J.}\ \bibnamefont {Yue}}, \bibinfo {author} {\bibfnamefont {W.~B.}\ \bibnamefont {Rui}}, \bibinfo {author} {\bibfnamefont {X.-M.}\ \bibnamefont {Zhu}}, \bibinfo {author} {\bibfnamefont {S.~A.}\ \bibnamefont {Yang}},\ and\ \bibinfo {author} {\bibfnamefont {Y.~X.}\ \bibnamefont {Zhao}},\ }\bibfield  {title} {\bibinfo {title} {Symmetry determined topology from flux dimerization},\ }\href {https://doi.org/10.1103/PhysRevB.109.115155} {\bibfield  {journal} {\bibinfo  {journal} {Phys. Rev. B}\ }\textbf {\bibinfo {volume} {109}},\ \bibinfo {pages} {115155} (\bibinfo {year} {2024})}\BibitemShut {NoStop}%
\end{thebibliography}

%

\end{document}